\documentclass[sn-mathphys-ay]{sn-jnl}

\usepackage{graphicx}%
\usepackage{multirow}%
\usepackage{amsmath,amssymb,amsfonts}%
\usepackage{amsthm}%
\usepackage{mathrsfs}%
\usepackage[title]{appendix}%
\usepackage{xcolor}%
\usepackage{textcomp}%
\usepackage{manyfoot}%
\usepackage{booktabs}%
\usepackage{algorithm}%
\usepackage{algorithmicx}%
\usepackage{algpseudocode}%
\usepackage{listings}%

\theoremstyle{thmstyleone}%
\theoremstyle{thmstyletwo}%

\theoremstyle{thmstylethree}%

\begin{document}

\title[Analytical theory of the spin-orbit state of a binary asteroid deflected by a kinetic impactor]{Analytical theory of the spin-orbit state of a binary asteroid deflected by a kinetic impactor}


\author*[1]{\fnm{Michalis} \sur{Gaitanas}}\email{mgaitana@physics.auth.gr}

\author[2]{\fnm{Christos} \sur{Efthymiopoulos}}\email{cefthym@math.unipd.it}

\author[1]{\fnm{Ioannis} \sur{Gkolias}}\email{igkoli@physics.auth.gr}

\author[1]{\fnm{George} \sur{Voyatzis}}\email{voyatzis@auth.gr}

\author[1]{\fnm{Kleomenis} \sur{Tsiganis}}\email{tsiganis@auth.gr}

\affil[1]{\orgdiv{Department of Physics}, \orgname{Aristotle University of Thessaloniki}, \orgaddress{\city{Thessaloniki}, \postcode{54124}, \country{Greece}}}

\affil[2]{\orgdiv{Department of Mathematics}, \orgname{Universit\`a degli Studi di Padova}, \orgaddress{\street{Via Trieste 63}, \city{Padova}, \postcode{35121},  \country{Italy}}}


\abstract{We study the perturbed-from-synchronous librational state of a double asteroid, modeled by the Full Two Rigid Body Problem (F2RBP), with primary emphasis on deriving analytical formulas which describe the system's evolution after deflection by a kinetic impactor. To this end, both a linear and nonlinear (canonical) theory are developed. We make the simplifying approximations (to be relaxed in a forthcoming paper) of planar binary orbit and axisymmetric shape of the primary body. To study the effect of a DART-like hit on the secondary body, the momentum transfer enhancement parameter $\beta$ is introduced and retained as a symbolic variable throughout all formulas derived, either by linear or nonlinear theory. Our approach can be of use in the context of the analysis of the post impact data from kinetic impactor missions, by providing a precise modeling of the impactor's effect on the seconadry's librational state as a function of $\beta$.}




\maketitle

\section{Introduction}\label{sec:intro}

The historic DART (Double Asteroid Redirection Test) experiment  was completed in September 2022, when the DART spacecraft impacted asteroid Dimorphos, the small secondary of (65803) Didymos, with the purpose of changing its 12-hour orbital period. Before impact, the system was assumed to be in a `relaxed state', i.e. a singly-synchronous configuration of the secondary on a circular orbit, around the swiftly rotating primary (see \cite{Richardson2022}). As expected, the DART impact induced a significant orbit period change, measured to $-33\pm 1$ minutes \citep{Thomas2023}. Simulations of the post-impact dynamics suggest that the momentum enhancement parameter $\beta$\footnote{The factor representing how many times the momentum of DART was in fact imparted to the target, given the momentum carried away from the system by the ejecta of the collision (see also Section \ref{subsec:postimpact}).} was even higher than expected, of order $\beta=3.6$ \citep{Cheng2023}. The imparted eccentricity on the new orbit was estimated to be of order 0.03 - later, \cite{Meyer2023} showed that the eccentricity may be slowly decreasing, a possible indication of shape deformation \citep{Raducan2024} and/or chaotic libration \citep{Agrusa2021}. It is interesting to note that the observed pre-impact shape, as derived in \cite{Daly2023} was much more oblate than originally thought, a factor that may play a critical role in the post-impact evolution of such systems.

In this paper, motivated by the DART experiment, we delve into the dynamics of a double asteroid system, perturbed from the exact single-synchronous state due to collision with a DART-like kinetic impactor. Our study is in the framework of the Full Two Rigid Body Problem (F2RBP) \citep{Macie1996}. We model analytically the post-impact librational state by two types of perturbation theories, hereafter referred to as i) linear, and ii) nonlinear canonical. In the former, the normal modes of a linearized system of equations of motion around a suitably defined equilibrium (representing the new synchronous state after the impact) are computed. In the latter, we use the Lie series method to compute series terms of order higher than linear in the small parameters of the problem.

To set clear the main steps in the formulation of our theories, we presently make two simplifying assumptions, to be relaxed in a forthcoming paper. The first is planar motion, restricting the two rigid bodies to move in the $x-y$ plane and to rotate only around their $z-$axis. This simplification reduces the problem's number of degrees of freedom, making the theory more manageable in terms of algebraic manipulations. The methods used, however, are straightforward to generalise in the full 3D problem, while the results found with the planar approach are already of practical utility in the applications. Secondly, we approximate the primary body as axisymmetric. As discussed in Section \ref{sec:probstate}, such an approximation is justified whenever the primary rotates rapidly with respect to the orbital motion (and rotation of the synchronous secondary), as for example in the case of the single synchronous equilibrium state assumed to have been holding in the case of 65803 Didymos before DART's impact. Under the above assumptions, we then show how to derive analytical functions of time, which describe the binary's evolution, retaining as a free parameter in all formulas the momentum transfer enhancement $\beta$, which plays a crucial role in the timeseries of the observable quantities to be recovered after the mission.

Over the years, noteworthy advancements have been made in both the analytical and numerical study of the dynamics of binary asteroids. Notable contributions include \citep{Kinoshita1972} who studied a binary system composed of a spherical and a triaxial body. \cite{Boue2009} studied the secular problem, using potential expansion up to order four and performed averaging over the fast angles. \cite{Scheeres2009} explored the stability of the F2RBP under the planar assumption of two triaxial bodies. \cite{McMahon2013} incorporated the axisymmetry of the primary body in the planar case and studied the bounds of librational motion of the secondary body. All these analytical results are obtained by the hypothesis of regular orbits. In the case of chaotic orbits, one has to rely on numerical methods. The numerical integrator GUBAS \citep{Hou2017,Davis2020,Agrusa2021} has been widely employed over the years for the purpose of numerically investigating the F2RBP. In one of these studies, \cite{Meyer2021} showed that potentially observable, post-impact orbit period variations, whose characteristics depend on the physical parameters of the system,  are induced to the mutual orbit. 

While most previous analytical studies of the librational regime of binary asteroids are based on some form of linear theory around the synchronous state, in the present paper we discuss a higher precision, nonlinear analytical theory, capable of addressing even complex waveforms of the times series of the various observable quantities corresponding to the librational state after a DART-like hit. A key element of our approach is the employment of a `book-keeping' method, which, at each order, collects in one (book-keeping) symbol all the various small quantities of the problem under study, grouped at similar orders of `smallness' (see Section \ref{sec:canonical}). Such small quantities are the eccentricity of the post-impact orbit, the amplitude of libration of the secondary, the effective $J_2$ term of the primary, the asphericity of the secondary, etc. We then show how, through the use of a unique book-keeping parameter, one can arrive easily at a suitable Birkhoff normalization process, by which a `normal form' i.e., a solvable Hamiltonian is constructed. 

In comparison, we examine also classical linear theory based on the representation of all motions as a superposition of two linear normal modes. Of these modes, one represents a torque-induced libration combined with a nearly circular orbit, while the other an eccentricity-induced libration combined with the corresponding eccentric orbit. Our nonlinear theory then describes the coupling between these two fundamental modes. Overall, examining both nonlinear and linear approaches gives us the chance to compare the accuracy of analytical theories with respect to numerical simulations, as obtained, e.g. by GUBAS.

The structure of the paper is as follows: Section \ref{sec:probstate} defines the problem in its general form and presents the two reductions mentioned earlier: planar motion and axisymmetric form of the primary body. We start with the definition of the initial conditions associated with the single synchronous equilibrium state, which serve as the point of departure for our study. Finally, we give some basic formulas, as well as a basic description of the effect on the system by the kinetic impactor, illustrated with numerical examples. Section \ref{sec:linear} describes our linear perturbation theory. The problem's normal modes are derived in their generic form. Then we discuss how to reconfigure the final solution as a function of the $\beta$ parameter. Section \ref{sec:canonical} introduces our canonical (normal form) approach. We expand the Hamiltonian into series in the above mentioned book-keeping parameter up to a desired order. Through the application of Birkhoff normalization by Lie series, we then construct a Hamiltonian normal form by which the analytical solution to the problem is obtained. Again, the numerical results are compared with those of the canonical theory. Section \ref{sec:conclusions} gives a summary of our conclusions.

\section{Problem statement}\label{sec:probstate}

\subsection{Hamiltonian}\label{subsec:ham}

\begin{figure}[h]
    \centering
	\includegraphics[scale=0.46]{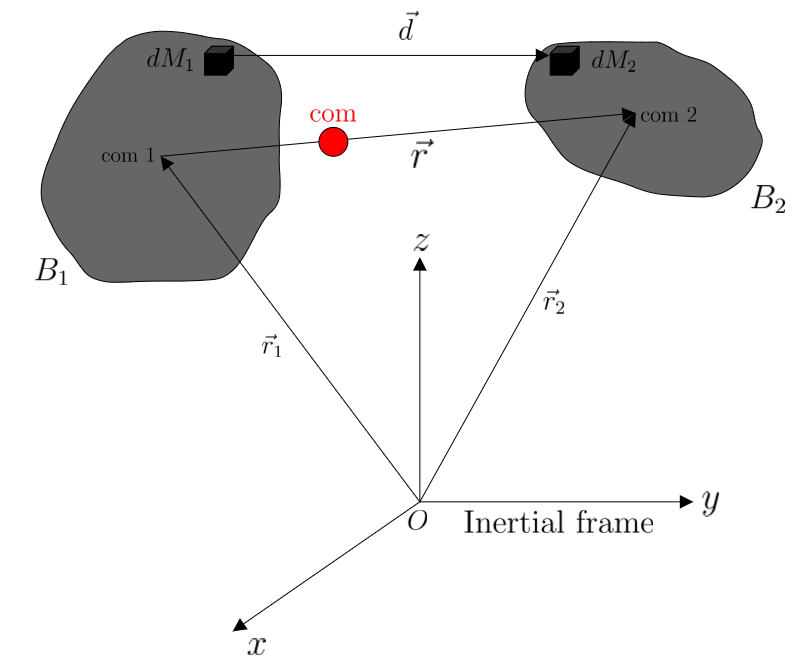}
    \caption{Definition of the F2RBP in the generic 3-dimensional form.}
    \label{fig1:f2bp3D}
\end{figure}

Let $B_1$ and $B_2$ be two rigid bodies of arbitrary mass distributions interacting gravitationally, embedded in a global inertial frame $Oxyz$ as in Figure \ref{fig1:f2bp3D}.

\newpage

\noindent
The generic form of the mutual potential function is

\begin{equation}\label{Voriginal}
    V = -\int\limits_{B_1}\int\limits_{B_2} \frac{G dM_1 dM_2}{d}~~~,
\end{equation}

\noindent
where $G$ is the gravitational constant. The integration takes place over both mass distributions, and $d$ is the distance between the finite mass elements $dM_1$ of $B_1$ and $dM_2$ of $B_2$. Equation (\ref{Voriginal}) can be expanded in the moments of inertia and up to the second order it reads \citep{Scheeres2009}

\begin{equation}\label{Vord2}
\begin{split}
V(\vec{r}, \boldsymbol{A}_1, \boldsymbol{A}_2) & = -\frac{GM_1M_2}{r} \\
      & - \frac{G}{2 r^3}
      \big[M_2 \text{tr}(\boldsymbol{\mathrm{I}}_1) + M_1 \text{tr}(\boldsymbol{\mathrm{I}}_2)\big] \\
      & + \frac{3G}{2r^5} \vec{r}\cdot(M_2 \boldsymbol{A}_1 \boldsymbol{\mathrm{I}}_1 \boldsymbol{A}_1^T + 
                                     M_1 \boldsymbol{A}_2 \boldsymbol{\mathrm{I}}_2 \boldsymbol{A}_2^T)
                                     \cdot \vec{r}~~~,
\end{split}
\end{equation}

\noindent
where $M_1,M_2$ are the primary's and secondary's masses, $\vec{r}$ the relative position vector between the two centers of mass, $\boldsymbol{A}_1, \boldsymbol{A}_2$ the rotation matrices of $B_1, B_2$ and $\boldsymbol{\mathrm{I}}_1, \boldsymbol{\mathrm{I}}_2$ the (constant) inertia matrices expressed in each body frame, centered around the corresponding center of mass. The system admits the energy and angular momentum integrals given by

\begin{equation}\label{EIntegral}
E = T + V = \frac{1}{2}\frac{M_1M_2}{M_1 + M_2}\vec{\upsilon}\cdot\vec{\upsilon} +
      \frac{1}{2}\vec{\omega}_1 \cdot \boldsymbol{\mathrm{I}}_1 \cdot \vec{\omega}_1 +
      \frac{1}{2}\vec{\omega}_2 \cdot \boldsymbol{\mathrm{I}}_2 \cdot \vec{\omega}_2 +
      V(\vec{r}, \boldsymbol{A}_1, \boldsymbol{A}_2)~~~,
\end{equation}

\begin{equation}\label{LIntegral}
\vec{L} = \frac{M_1M_2}{M_1 + M_2}\vec{r}\times\vec{\upsilon} +
          \boldsymbol{A}_1 \cdot (\boldsymbol{\mathrm{I}}_1 \cdot \vec{\omega}_1) +
          \boldsymbol{A}_2 \cdot (\boldsymbol{\mathrm{I}}_2 \cdot \vec{\omega}_2)~~~,
\end{equation}

\noindent
where $\vec{\upsilon}$ is the relative velocity vector between the two centers of mass and $\vec{\omega}_1, \vec{\omega}_2$ are the body frame angular velocity vectors of $B_1, B_2$.

We now introduce two approximations:

\noindent
1) \textit{Reduction to planar motion}: assuming the mutual orbit to be restricted to the $x-y$ plane and the rotation of both bodies to be restricted around their axes normal to the plane, the mutual potential function becomes

\begin{equation}\label{Vord2Plan}
\begin{aligned}
   V(r,\phi_1,\phi_2) = & -\frac{G M_1 M_2}{r} + 
   \frac{G M_2 ( I_{1x} + I_{1y} - 2I_{1z}) + G M_1(I_{2x} + I_{2y} - 2I_{2z}) }{4 r^3} \\
 & + \frac{3 G M_2  (I_{1x} - I_{1y}) \cos{(2 \phi_1)} }{4 r^3} +
   \frac{3 G M_1  (I_{2x} - I_{2y}) \cos{(2 \phi_2)} }{4 r^3}~~~.
\end{aligned}
\end{equation}

\noindent
The corresponding Hamiltonian is

\begin{equation}\label{Hord2Plan}
H = T + V =
\frac{p_r^2}{2m} +
\frac{p_{\phi_1}^2}{2I_{1z}} +
\frac{p_{\phi_2}^2}{2I_{2z}} +
\frac{ (p_{\theta} - p_{\phi_1} - p_{\phi_2})^2 }{2mr^2} +
V(r,\phi_1,\phi_2)~~~.
\end{equation}

\begin{figure}[h]
    \centering
	\includegraphics[scale=0.38]{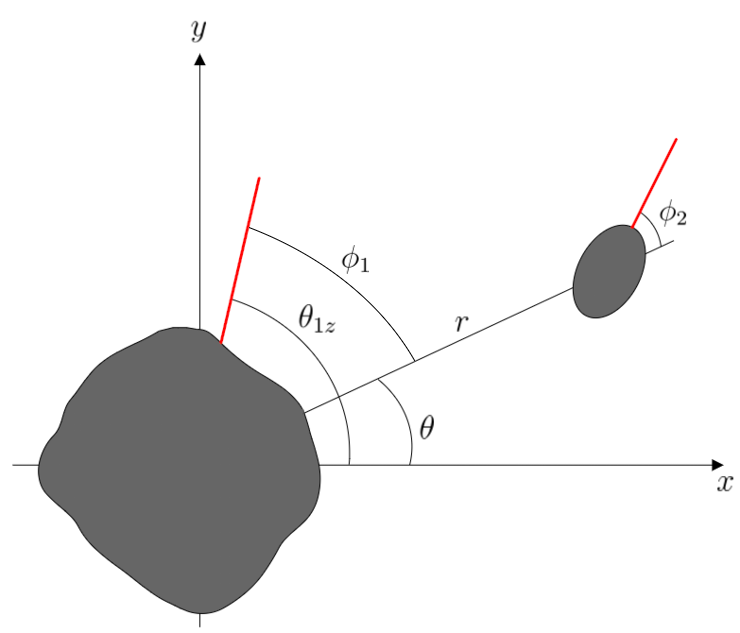}
    \caption{Schematic representation of the planar motion constraint. The two bodies are restricted to move in the $x-y$ plane and to rotate only around their $z-\text{axis}$.}
    \label{fig2:f2bpxy}
\end{figure}

\noindent
Here, $m = (M_1 M_2)/(M_1+M_2)$ is the reduced mass of the system and $\phi_1, \phi_2$ are the relative yaw angles with respect to the orbital displacement angle $\theta$. We refer below also to the angle $\theta_{1z} = \theta + \phi_1$ (see Figure \ref{fig2:f2bpxy}). Finally, $p_r, p_\theta, p_{\phi_1}, p_{\phi_2}$ are the canonical momenta, conjugate to the coordinates $r, \theta, \phi_1, \phi_2$. We have $p_r = m\dot{r}$, $p_{\phi_1} = I_{1z}(\dot{\theta} + \dot{\phi}_1)$, $p_{\phi_2} = I_{2z}(\dot{\theta} + \dot{\phi}_2)$, while $p_\theta = mr^2\dot{\theta} + p_{\phi_1} + p_{\phi_2}$ which is the total angular momentum of the system.

\noindent
2) \textit{Approximation of the primary body as axisymmetric}: such an approximation is justified whenever the primary is a fast rotator, so that all its non-axisymmetric multipole moments lead to fast oscillations which can be eliminated from the equations of motion. Formally, this implies adopting as value for both $I_{1x}, I_{1y}$ the mean of the real values, $I_s = (I_{1x}+I_{1y})/2$. The latter assumption eliminates the Hamiltonian's dependence on the angle $\phi_1$ (or $\theta_{1z}$), thus giving rise to four equilibrium points, corresponding to the values $\phi_2 = 0, \pi, \pm \pi/2$. The final Hamiltonian then takes the form 

\begin{equation}\label{HJ2ell}
H = \frac{p_r^2}{2m} +
\frac{p_{\phi_1}^2}{2I_{1z}} +
\frac{p_{\phi_2}^2}{2I_{2z}} +
\frac{ (p_{\theta} - p_{\phi_1} - p_{\phi_2})^2 }{2mr^2} +
V(r,\phi_2)~~~,
\end{equation}

\noindent
where

\begin{equation}\label{VJ2ell}
\begin{split}
V(r,\phi_2) & = -\frac{G M_1 M_2}{r} + 
                 \frac{G M_1 ( I_{2x} + I_{2y} - 2I_{2z}) + 2GM_2(I_s - I_{1z}) }{4 r^3}  \\
                 & + \frac{3GM_1 (I_{2x} - I_{2y}) \cos{(2\phi_2)} }{4 r^3}~~~.
\end{split}
\end{equation}

\subsection{Pre-impact synchronous equilibrium state}\label{subsec:preimpact}

\noindent
Given the above approximations, and assuming a (fast) rotation of the primary with angular velocity $\nu_1 = \dot{\theta}_{1z}(0)$, the following set of initial conditions define the so-called single synchronous equilibrium state of the binary at a separation (distance between the centers of mass) equal to $r_{eq}$:

\begin{figure}[H]
    \centering
	\includegraphics[scale=0.47]{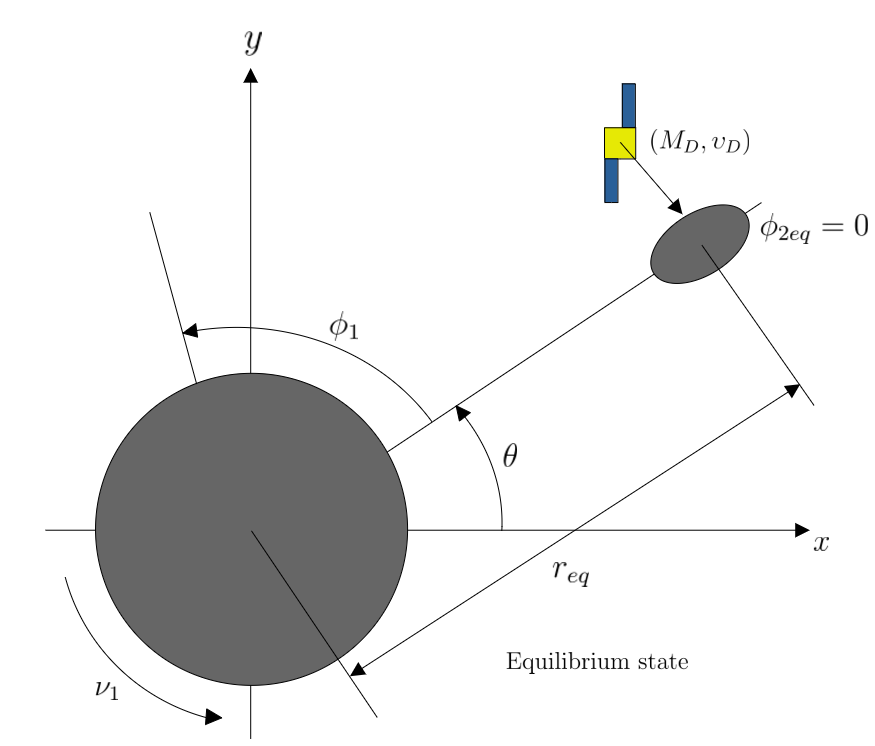}
    \caption{Representation of the pre-impact planar equilibrium configuration.}
    \label{fig3:f2bphit}
\end{figure}

\begin{equation}\label{J2ellEquilPos}
\begin{gathered}
   r(0) = r_{eq} \\
   \phi_2(0) = \phi_{2eq} = 0
\end{gathered} 
\end{equation}
\vspace{0.1cm}

\begin{equation}\label{J2ellEquilVel}
\begin{gathered}
  \dot{r}(0) = \dot{r}_{eq} = 0 \\
  \dot{\theta}(0) = \dot{\theta}_{eq} = \sqrt{ 
    \frac{G(M_1 + M_2)}{r_{eq}^3}
    \left[ 1 + \frac{3}{2 r_{eq}^2}
            \left( \frac{I_{1z} - I_s}{M_1} + \frac{I_{2y} + I_{2z} - 2I_{2x}}{M_2} \right) 
    \right] } \\
    \dot{\phi}_1(0) = \dot{\phi}_{1eq} = \nu_1 - \dot{\theta}_{eq} \\
  \dot{\phi}_2(0) = \dot{\phi}_{2eq} = 0
\end{gathered} 
\end{equation}
\vspace{0.1cm}

\begin{equation}\label{J2ellEquilMom}
\begin{gathered}
 p_r(0) = p_{r_{eq}} = 0 \\
 p_{\theta}(0) = p_{\theta_{eq}} = p_{\phi_{1eq}} + p_{\phi_{2eq}} + mr_{eq}^2 \dot{\theta}_{eq} \\
 p_{\phi_1}(0) = p_{\phi_{1eq}} = \nu_1I_{1z} \\
 p_{\phi_2}(0) = p_{\phi_{2eq}} = \dot{\theta}_{eq}I_{2z}~~~.
\end{gathered} 
\end{equation}
\vspace{0.1cm}

\noindent
We note that the entire set of the above initial conditions are determined by two free parameters, namely $r_{eq}$ and $\nu_1$ (see Figure \ref{fig3:f2bphit}).

\subsection{Post-impact evolution}\label{subsec:postimpact}

\noindent
Assume, now, that the secondary undergoes a DART-like head-on collision, with the impactor's velocity vector lying always in the $x-y$ plane, being opposite to the velocity vector of the secondary, and pointing towards the secondary's center (Figure \ref{fig3:f2bphit}). If $M_D, \upsilon_D$ are the mass and velocity magnitude of the impactor, then the velocity change $\Delta \upsilon$ induced to the secondary can be expressed in terms of the \textit{momentum enhancement factor $\beta$} through the equation \citep{Rivkin2021}

\begin{equation}
\Delta \upsilon = -\frac{\beta M_D\upsilon_D}{M_2}~~~.
\end{equation}

\noindent
Because the recoiled ejecta are expelled in the opposite direction of the impactor's velocity, the momentum change is enhanced, yielding $\beta > 1$. If no ejecta are expelled, the collision is considered perfectly inelastic, resulting in $\beta = 1$.

Depending on the parameters of the impactor and of the binary, we distinguish two possible evolution cases after the impact: regular or chaotic. In a chaotic case, little can be done in terms of analytical theories as indicated in some cases found by \cite{Agrusa2021} and hence, one must resort to numerical integration. If the collision is not `violent' enough to provoke chaotic evolution, the post impact state remains regular and can be described approximately by analytical formulas, which we focus on the present paper. Physically, after the impact, the trajectory turns from circular to eccentric, with the apocenter initially coinciding with the point of impact, while the secondary also develops a libration $\phi_2(t) \neq 0$. 

\begin{figure}[H]
  \centering
    \includegraphics[scale=0.7]{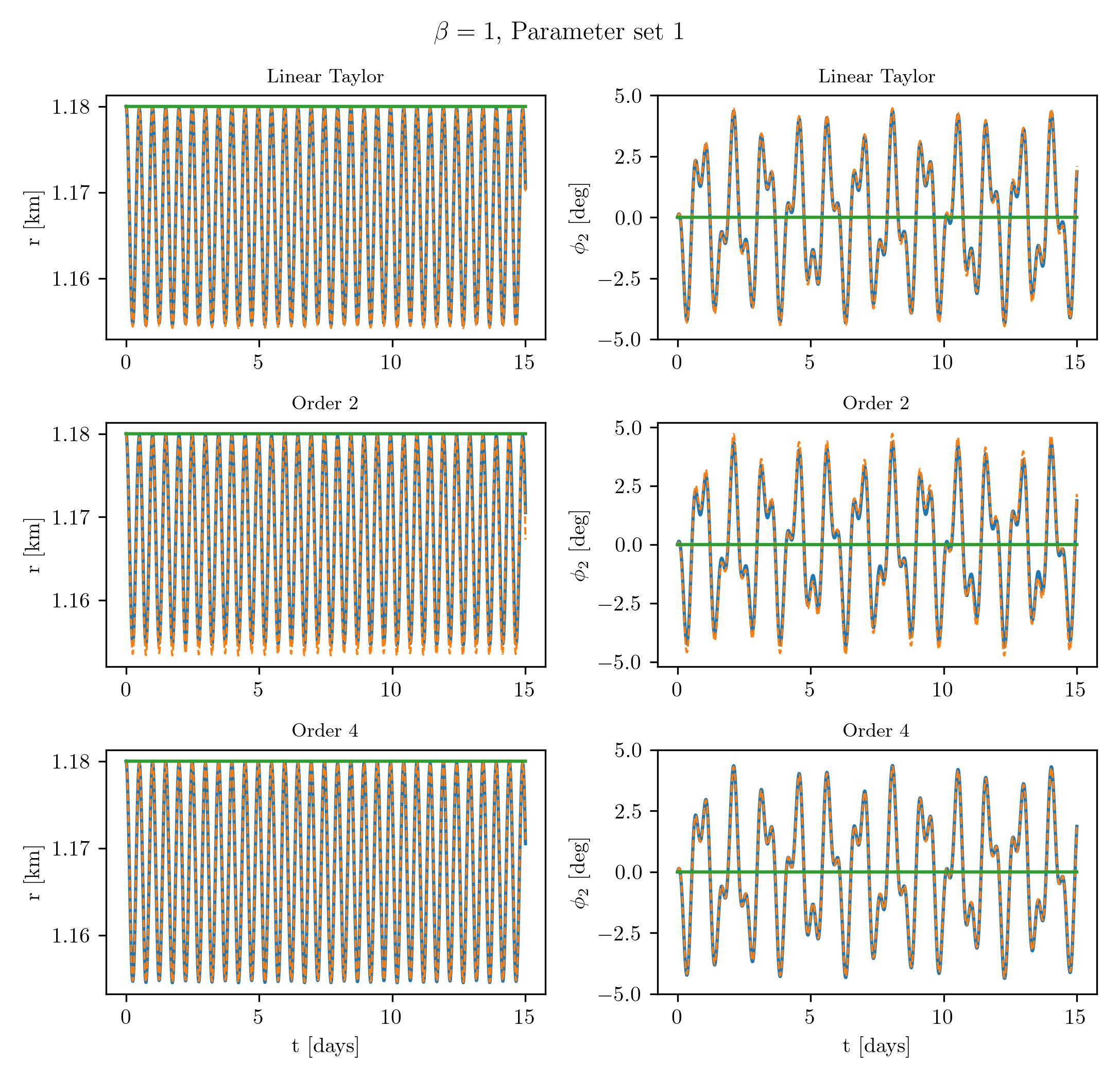}
    \caption{Post impact evolution of the functions $r(t)$ (left column) and $\phi_2(t)$ (right column) for
    the case $\beta = 1$ and for the first set of physical parameters
    ($G = 0.0864989$, $M_1 = 5.15045$, $M_2 = 0.0392647$,
      $I_s = 0.263844$, $I_{1z} = 0.337921$, $I_{2x} = 8.20357 \cdot 10^{-5}$,
      $I_{2y} = 8.88678 \cdot 10^{-5}$, $I_{2z} = 1.18976 \cdot 10^{-4}$,
      $r_{eq} = 1.18$, $\nu_1 = 2\pi/2.26$,
      $M_D = 5.79434 \cdot 10^{-9}$, $\upsilon_D = 22121.6$). 
      The units of measurement are $[hr], [km], [kg^{\ast}]$, where $1 [kg^{\ast}] = 10^{11}[kg]$. The blue curves correspond to the results derived via numerical integration, while the dashed orange curves correspond to the results obtained via each of the examined analytical theory i.e., as indicated on top of each panel. The green curves represent the initial (pre-impact) equilibrium state of the binary.}
      \label{fig4:set1b1}
\end{figure}

\begin{figure}[H]
  \centering
    \includegraphics[scale=0.7]{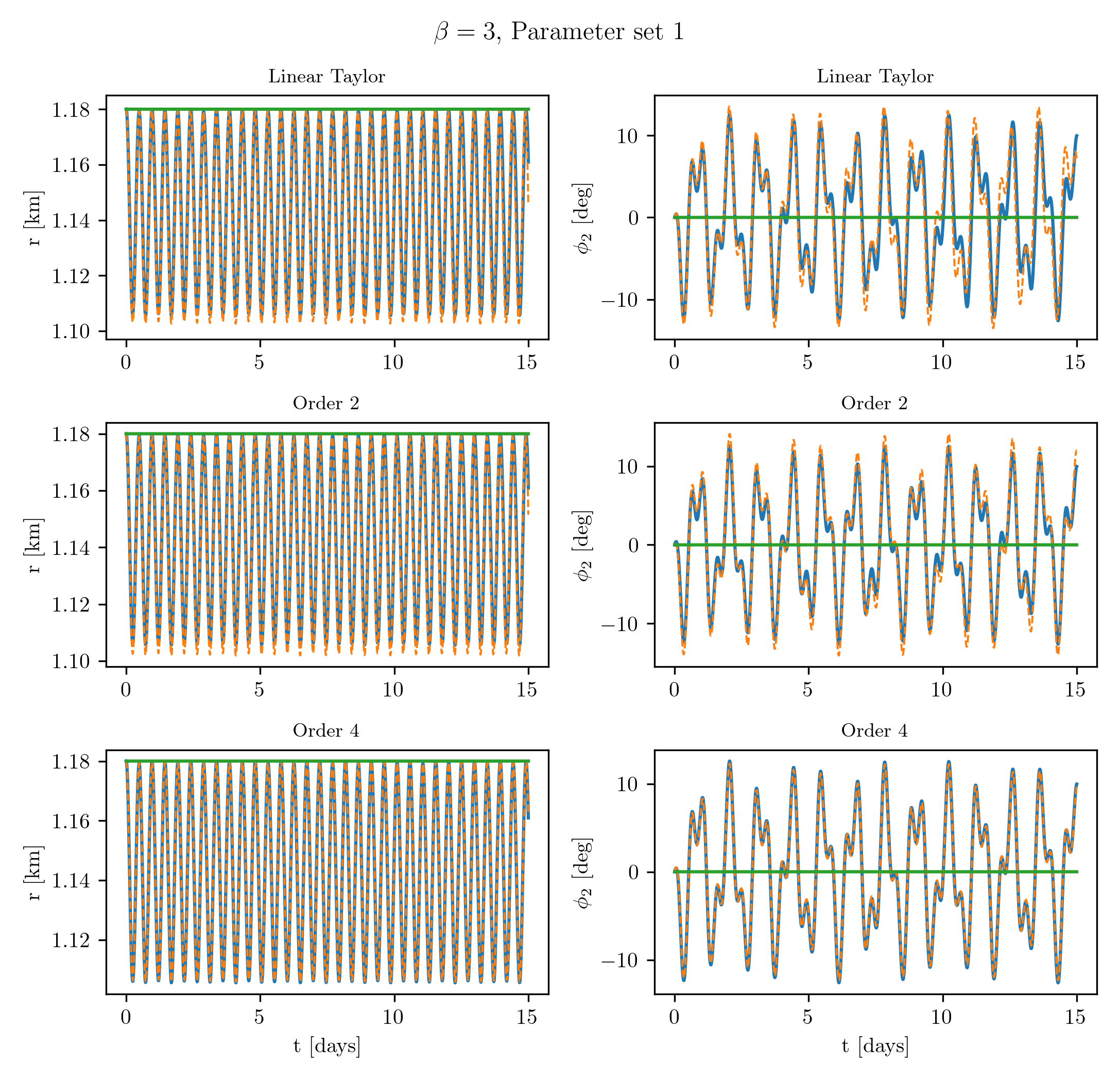}
    \caption{Same as in Figure \ref{fig4:set1b1}, but for the case $\beta = 3$.}
    \label{fig5:set1b3}
\end{figure}

\begin{figure}[H]
  \centering
    \includegraphics[scale=0.7]{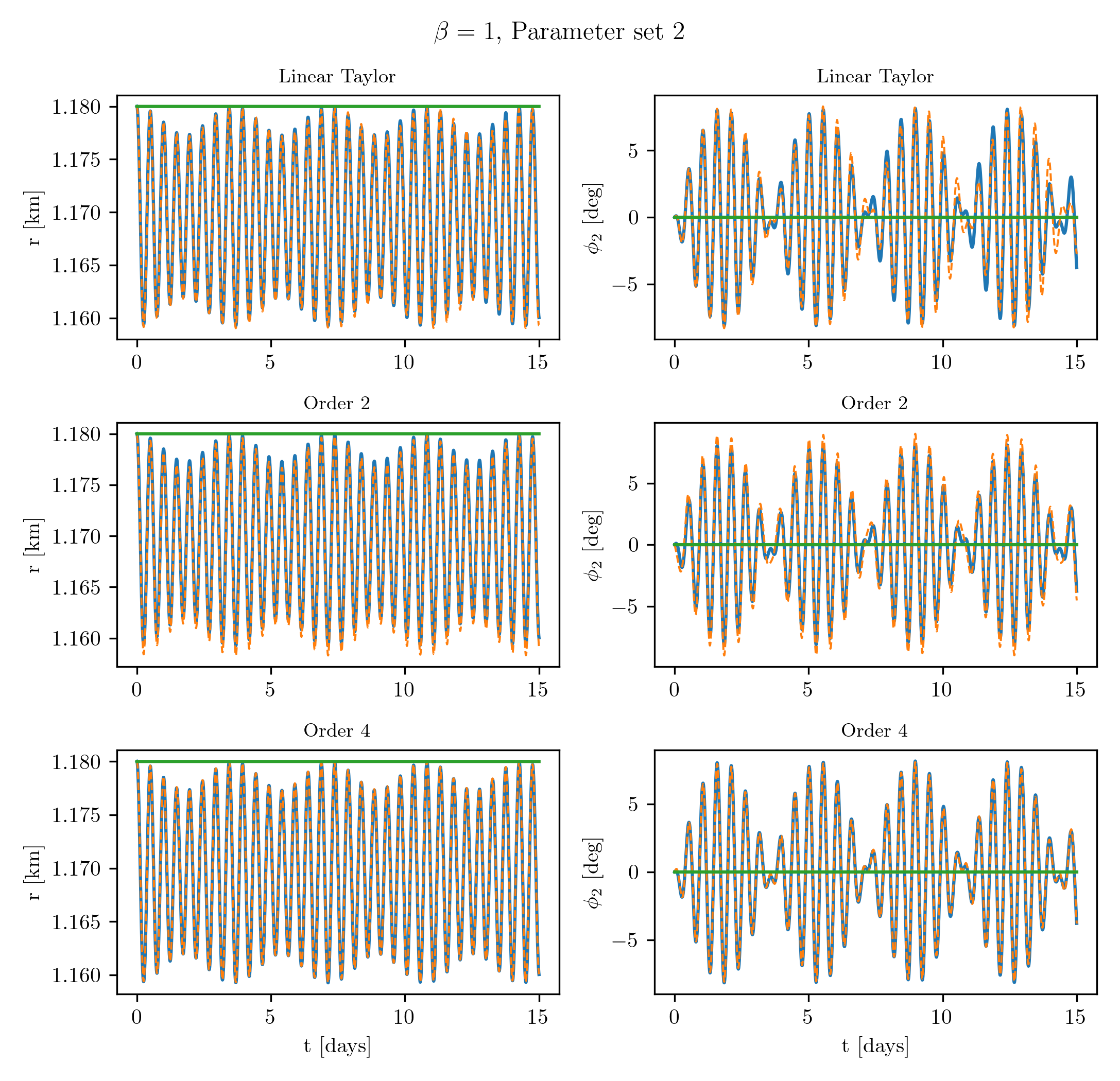}
    \caption{Same as in Figure \ref{fig4:set1b1}, but for the case $\beta = 1$ and for the second set of physical parameters
    ($G = 0.0864989$, $M_1 = 5.14073$, $M_2 = 0.048272$,
      $I_s = 0.254128$, $I_{1z} = 0.326872$, $I_{2x} = 9.28891 \cdot 10^{-5}$,
      $I_{2y} = 1.27131 \cdot 10^{-4}$, $I_{2z} = 1.29876 \cdot 10^{-4}$),
      $r_{eq} = 1.18$, $\nu_1 = 2\pi/2.26$,
      $M_D = 5.79434 \cdot 10^{-9}$, $\upsilon_D = 22121.6$). The units remain the same.}
      \label{fig6:set2b1}
\end{figure}

\begin{figure}[H]
  \centering
    \includegraphics[scale=0.7]{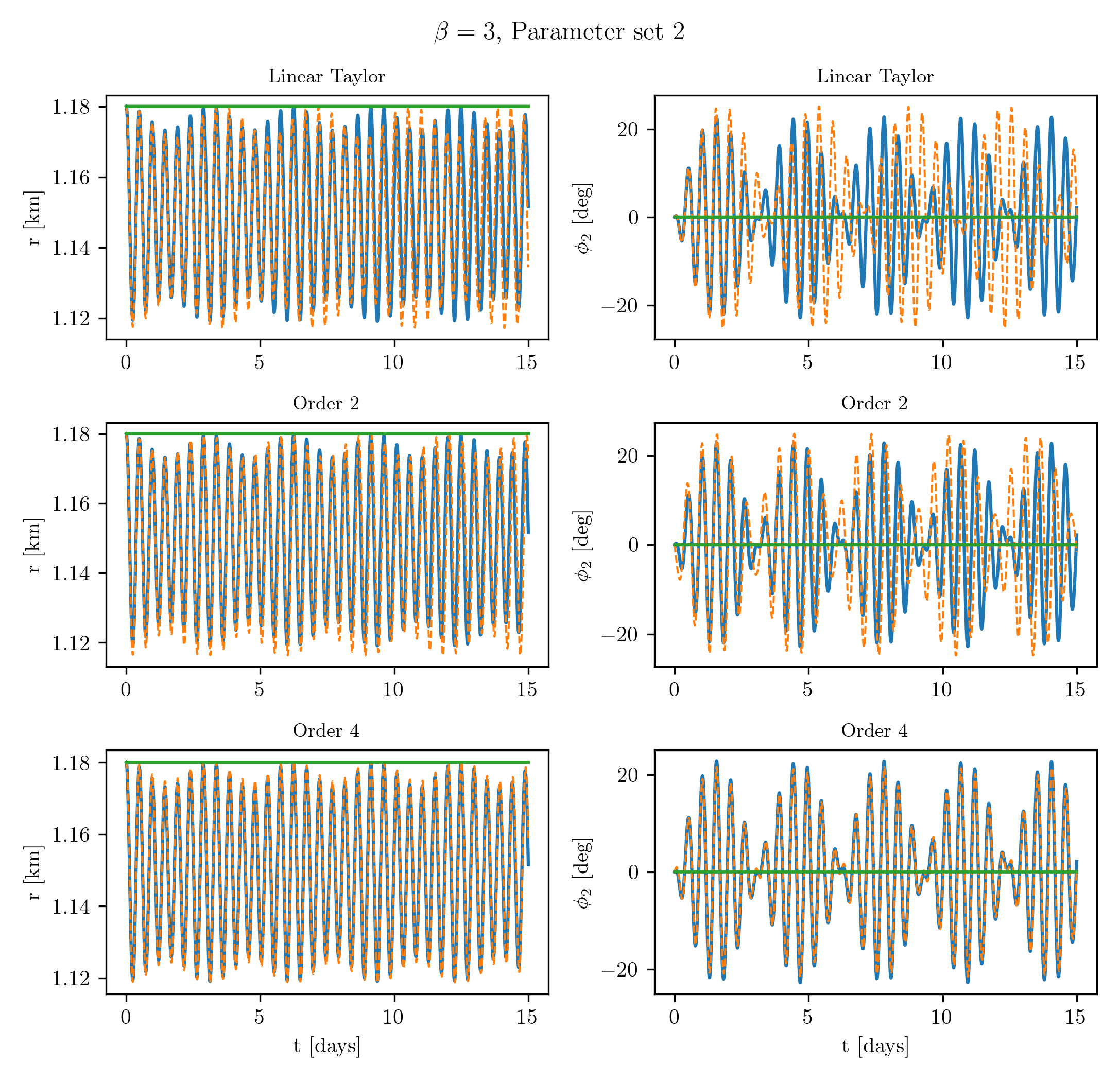}
    \caption{Same as in Figure \ref{fig6:set2b1}, but for the case $\beta = 3$.}
    \label{fig7:set2b3}
\end{figure}

Figures \ref{fig4:set1b1} to \ref{fig7:set2b3} show four examples of the numerical evolution of the functions $r(t)$ and $\phi_2(t)$ post impact (blue curves), obtained by assuming two different sets of physical parameters for the binary system, as well as two possible values of $\beta$ for each set. Typically, we find that $r(t)$ just follows a low amplitude epicyclic oscillation due to the eccentricity induced to the orbit. On the other hand, the angle $\phi_2(t)$ starts exhibiting a librational behavior. As shown in the next section, this is dominated by two harmonics, corresponding to two fundamental frequencies, and several linear combinations thereof. In general, the amplitude of the libration increases with $\beta$ and it can reach several degrees, leading to a substantially nonlinear behavior of the system. In the case of regular motion, nonlinearity manifests itself mostly in the fact that the fundamental frequencies of libration, as derived by a numerical analysis, differ from the corresponding linear (normal mode) frequencies. This fact causes a slow dephasing of the analytical solutions from the numerical ones, as eye-evident, for example, in Figure \ref{fig7:set2b3}. The orange-dashed curves in each panel of Figures \ref{fig4:set1b1} to \ref{fig7:set2b3} show the corresponding analytical prediction for each orbit, obtained by the different theories discussed in detail in Sections \ref{sec:linear} and \ref{sec:canonical} below. While deferring a detailed discussion to the corresponding Sections, we observe, already at this stage, that the reduction of the error of analytical theories at higher order is due, mostly, to the improvement in the predictions by which a theory recovers the numerical values of the fundamental frequencies as the order of the theory increases, preserving the parameter $\beta$ at the same time. We now turn our attention to the analytical methods used to achieve such improvement.

\section{Linear perturbation theory}\label{sec:linear}

\subsection{New equilibrium}\label{subsec:neweq}

\noindent
As soon as the impact occurs, the binary is deflected away from its pre-impact synchronous equilibrium state. Formally, the momentum transfer by the impactor affects the value of the integral of motion $p_\theta$ in the Hamiltonian (\ref{HJ2ell}), a head-on collision leading to a decrease of $p_{\theta}$. In terms of the parameter $\beta$, the post-impact state then starts from the initial conditions

\begin{equation}\label{ImpactStateEquations}
\begin{gathered}
    \dot{r}_{imp}(0) = 0, \hspace{0.2cm} p_{r_{imp}}(0) = p_{r_{eq}} = 0 \\
    \dot{\theta}_{imp}(0) = \dot{\theta}_{eq} - \frac{M_D\upsilon_D\beta}{r_{eq} M_2}, \hspace{0.2cm}
    p_{\theta_{imp}}(0) = p_{\theta_{eq}} - \frac{mr_{eq}M_D\upsilon_D\beta}{M_2} \\
    \dot{\phi}_{1_{imp}}(0) = \nu_1 - \dot{\theta}_{imp}(0), \hspace{0.2cm} p_{\phi_{1imp}}(0) = p_{\phi_{1eq}} \\
    \dot{\phi}_{2_{imp}}(0) = \frac{M_D\upsilon_D\beta}{r_{eq} M_2}, \hspace{0.2cm} p_{\phi_{2imp}}(0) = p_{\phi_{2eq}}~~~.
\end{gathered}
\end{equation}

\noindent
Since the ratio $M_D/M_2$ in (\ref{ImpactStateEquations}) is a small quantity, the resulting new state can be studied as a small deviation from a new equilibrium state, corresponding to the new value of the integral $p_{\theta} = p_{\theta_{imp}}$. We specify the (constant in time) distance $r_{eq,new}$ of the two bodies at the new equilibrium by solving (for $r_{eq,new}$) the equation

\begin{equation}\label{NewEqState}
\begin{split}
 \dot{\theta}_{eq,new} & =
 \frac{(I_{2z} + mr_{eq}^2)\dot{\theta}_{eq}}{I_{2z} + mr^2_{eq,new}} -
 \frac{mr_{eq}M_D\upsilon_D\beta}{M_2(I_{2z} + mr^2_{eq,new})} \\
 & = \sqrt{ 
    \frac{G(M_1 + M_2)}{r_{eq,new}^3}
    \left[ 1 + \frac{3}{2 r_{eq,new}^2}
            \left( \frac{I_{1z} - I_s}{M_1} + \frac{I_{2y} + I_{2z} - 2I_{2x}}{M_2} \right) 
    \right] }~~~.
\end{split}
\end{equation}

\noindent
Note that, in order to keep $\beta$ as a symbol in all the expressions related to the new equilibrium state, one has to perform a Taylor expansion of equation (\ref{NewEqState}) in the small quantity $\delta r = r_{eq} - r_{eq,new}$ around the value $\delta r = 0$, which corresponds to no change in the kinetic state. We find

\begin{equation}\label{reqnewLin}
r_{eq,new} = r_{eq} + C_{r_{eq}}\beta + O(\delta r^2)~~~,
\end{equation}

\noindent
where

\begin{equation}\label{reqnewLinCoeff}
\begin{gathered}
    C_{r_{eq}} =
\frac{ 2mr_{eq}^{9/2}\sqrt{6C_I+4r_{eq}^2}M_D\upsilon_D }
     { M_2\sqrt{G(M_1+M_2)}\big( 6I_{2z}r_{eq}^2 - 2mr_{eq}^4 +3C_I(5I_{2z} + mr_{eq}^2) \big)} \\
    C_I = \frac{I_{1z} - I_s}{M_1} + \frac{-2I_{2x} + I_{2y} + I_{2z}}{M_2}~~~.
\end{gathered} 
\end{equation}

\begin{figure}[H]
  \centering
    \includegraphics[scale=0.52]{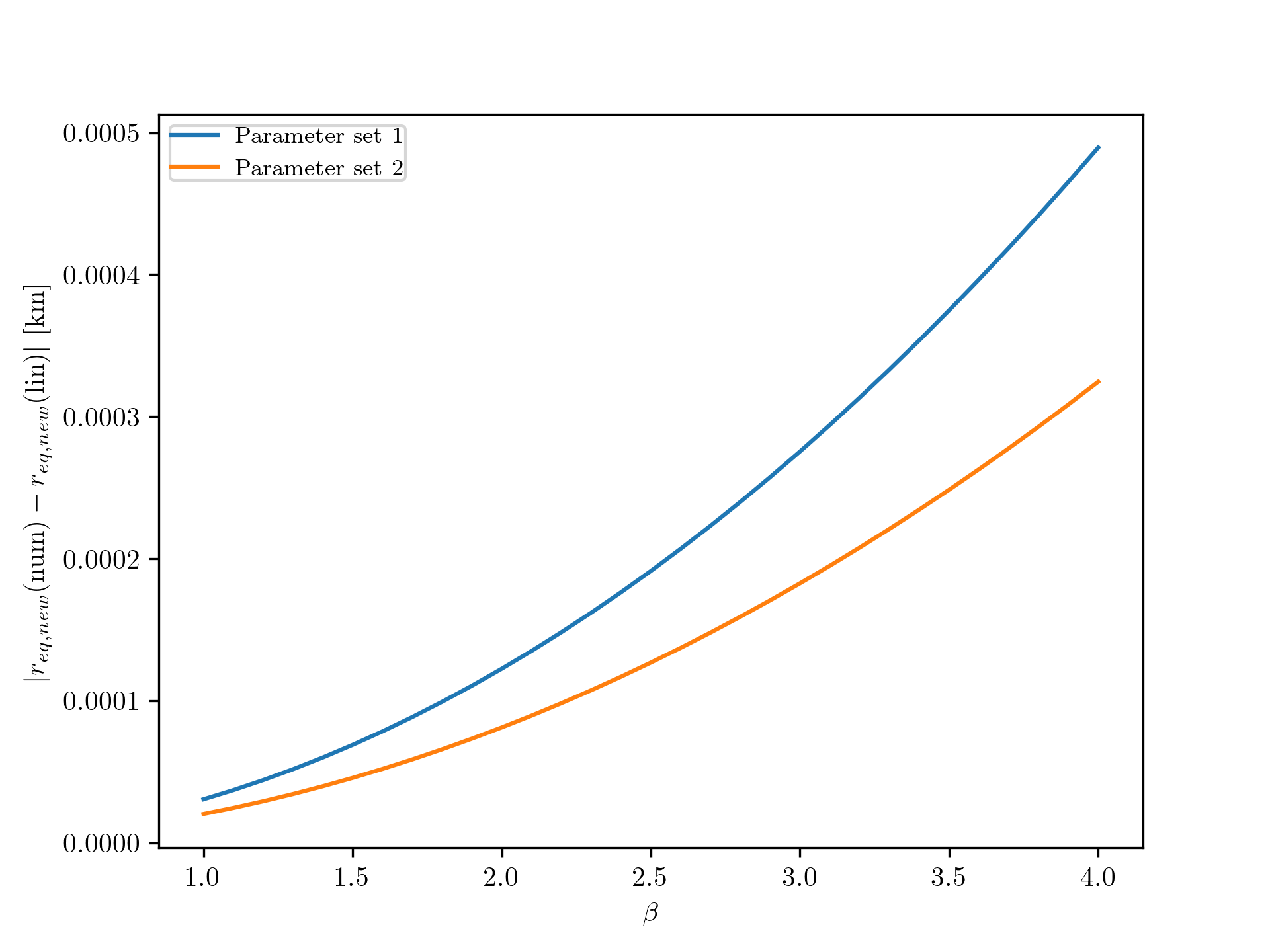}
    \caption{Absolute error between the new equilibrium point, computed via root-finding and
    the one computed via the `Linear Taylor' approximation.}
    \label{fig8:eqperror}
\end{figure}

\noindent
Note that even if $\beta > 1$, the quantity $C_{r_{eq}}\beta$ is $O(M_D/M_2)$, hence small. Substituting $r_{eq,new}$, as given by equation (\ref{reqnewLin}), whenever necessary in the formulas of the linear theory leads to a method referred to as `Linear Taylor' (see Figures \ref{fig4:set1b1} to \ref{fig7:set2b3}). Using a root-finding algorithm to numerically determine the exact solution of equation (\ref{NewEqState}) is also possible, but only marginally alters the final results. Figure \ref{fig8:eqperror} depicts the absolute error between the new equilibrium point obtained via root-finding and the one via equation (\ref{reqnewLin}). For the expected values of $\beta$, the maximum error is within the order of a few decimetres.

\subsection{Linear analytical solution for the post-impact state}\label{subsec:lineartheory}

\noindent
Since the angles $\theta$ and $\phi_1$ are ignorable in the Hamiltonian (\ref{HJ2ell}), the quantities $p_{\theta}$ and $p_{\phi_1}$ are constants of motion, whose values are determined only by the initial conditions, i.e., by equations (\ref{J2ellEquilPos}) - (\ref{J2ellEquilMom}), with $r_{eq}$ replaced by $r_{eq,new}$, found in turn through equation (\ref{reqnewLin}) or numerically. We then consider the 4x4 reduced system of Hamilton's equations in the variables $(r, \phi_2,  p_r, p_{\phi_2})$

\begin{equation}\label{J2EllEOMS}
\begin{gathered}
    \dot{r} = \frac{\partial H}{\partial p_r} \hspace{1.1cm} \dot{p}_r = -\frac{\partial H}{\partial r} \\
    \dot{\phi}_2 = \frac{\partial H}{\partial p_{\phi_2}} \hspace{1cm} \dot{p}_{\phi_2} = -\frac{\partial H}{\partial \phi_2}~~~,
\end{gathered} 
\end{equation}

\noindent
where $H = H(r,\phi_2, p_r, p_{\phi_2}; p_\theta = p_{\theta_{imp}}, p_{\phi_1} = p_{\phi_{1eq}})$. Setting $r = r_{eq,new} + \delta r$, \mbox{$\phi_2 = 0 + \delta \phi_2$}, $p_r = 0 + \delta p_r, p_{\phi_2} = p_{\phi_{2eq,new}} + \delta p_{\phi_2}$, with $p_{\phi_{2eq,new}} = \dot{\theta}_{eq,new}(r_{eq,new})I_{2z}$, and linearizing the 4x4 equations of motion around the new equilibrium point, leads to the linear system

\begin{equation}\label{LinSys}
\dot{\overrightarrow{\delta X}} = \boldsymbol{J} \cdot \overrightarrow{\delta X}~~~,
\end{equation}

\noindent
where

\begin{equation}\label{LinJac}
\overrightarrow{\delta X} = [\delta r, \delta \phi_2, \delta p_r, \delta p_{\phi_2}]^T \hspace{0.4cm} \text{and} \hspace{0.4cm}
\boldsymbol{J} =
\begin{bmatrix}
      0    &   0    & j_{13} &   0    \\
    j_{21} &   0    &   0    & j_{24} \\
    j_{31} &   0    &   0    & -j_{21} \\
      0    & j_{42} &   0    &   0
\end{bmatrix}~~~.
\end{equation}

\noindent
The Jacobian's non-zero elements $j_{nm}$ (evaluated at the new equilibrium point) depend only on the adopted values of the physical parameters of the binary and on the value of $\beta$ (through their dependence on $r_{eq,new}$ in equation (\ref{reqnewLin}). The characteristic polynomial is

\begin{equation}\label{PolJ2ell}
P(\lambda) = \lambda ^4 + \xi_2\lambda ^2 + \xi_0~~~,
\end{equation}

\noindent
where the analytical expressions for $j_{nm}$, $\xi_0$ and $\xi_2$ are given in the Appendices \ref{app:jnm} and \ref{app:xi02} respectively. The roots of the characteristic polynomial are $\lambda_{1,2} = \mp i\omega_1$, $\lambda_{3,4} = \mp i\omega_2$, where the two fundamental frequencies

\begin{equation}\label{J2EllOmegas}
    \omega_{1,2} = \frac{\sqrt{\xi_2 \pm \sqrt{\xi_2 ^2-4 \xi_0 } }}{\sqrt{2}}~~~,
\end{equation}

\noindent
correspond, respectively, to the normal modes associated with the eccentricity-induced and with the torque-induced librational motion of the secondary. General expressions for the solution of the system (\ref{LinSys}) are easy to compute for an arbitrary initial excitation $\big(\delta r(0), \delta \phi_2(0), \delta p_r(0), \delta p_{\phi_2}(0)\big)$, by superposing the solutions corresponding to the two normal modes. However, the resulting expressions are cumbersome and will not be presented here. We focus, instead, on the simple case of initial conditions corresponding to a DART-like hit, for which $\delta r(0) = r_{eq} - r_{eq,new}$, $\delta \phi_2(0) = 0$, $\delta p_r(0) = 0$, $\delta p_{\phi_2}(0) = p_{\phi_{2eq}} - p_{\phi_{2eq,new}}$ (Appendix \ref{app:dartlike}). We then obtain the expressions

\begin{align}
\centering   
\delta r(t) =& A_{1r}\cos{(\omega_1 t)} + A_{2r}\cos{(\omega_2 t)} \nonumber\\
\delta \phi_2(t) =&
\frac{\big( j_{13}(j_{21}^2 + j_{24}j_{31}) + j_{24}\omega_1^2 \big)A_{1r}}{j_{13}j_{21}\omega_1}\sin{(\omega_1 t)} \nonumber\\
&+ \frac{\big( j_{13}(j_{21}^2 + j_{24}j_{31}) + j_{24}\omega_2^2 \big)A_{2r}}{j_{13}j_{21}\omega_2}\sin{(\omega_2 t)} \label{LinVar} \\
\delta p_r(t) =&
-\frac{\omega_1A_{1r}}{j_{13}}\sin{(\omega_1 t)} - \frac{\omega_2A_{2r}}{j_{13}}\sin{(\omega_2 t)} \nonumber\\
\delta p_{\phi_2}(t) =&
\frac{(j_{13}j_{31} + \omega_1^2)A_{1r}}{j_{13}j_{21}}\cos{(\omega_1 t)} +
\frac{(j_{13}j_{31} + \omega_2^2)A_{2r}}{j_{13}j_{21}}\cos{(\omega_2 t)} \nonumber ~~~,
\end{align}

\noindent
where the coefficients $A_{1r}, A_{2r}$ are functions of $j_{nm}$, $\omega_1$, $\omega_2$, $\delta r(0)$ and $\delta p_{\phi_2}(0)$, and are given in the Appendix \ref{app:lincoeff}. Due to the explicit dependence of $r_{eq,new}$ on $\beta$ (see equation (\ref{reqnewLin})), it turns out that the linear amplitudes, frequencies and hence, the whole linear solution can be written using explicit formulas as functions of $\beta$. We then obtain an analytical (called `Linear-Taylor') approximation $\big(r_{lin}(t;\beta), \phi_{2lin}(t;\beta), p_{r_{lin}}(t;\beta), p_{\phi_{2lin}}(t;\beta) \big)$ of the solution of Hamilton's equations for a DART-like hit given by

\begin{equation}\label{LinSol}
\begin{gathered}   
    r_{lin}(t;\beta) = r_{eq,new}(\beta) + \delta r(t;\beta) \\
    \phi_{2_{lin}}(t;\beta) = \delta \phi_2(t;\beta) \\
    p_{r_{lin}}(t;\beta) = \delta p_r(t;\beta) \\
    p_{\phi_{2lin}}(t;\beta) = p_{\phi_{2eq,new}}(r_{eq,new}(\beta)) + \delta p_{\phi_2}(t;\beta)~~~.
\end{gathered} 
\end{equation}
\vspace{0.1cm}

\noindent
Regarding the relative orbital displacement $\theta$, and the primary's relative rotation $\phi_1$, explicit functions of time and $\beta$ can be derived for both, assuming linear approximation to the right hand side of $\dot{\theta}$ (detailed derivation is presented in Appendix \ref{app:evalthetaphi}). We find

\begin{equation}\label{PapThetat}
\begin{split}
    \theta(t;\beta) & = \theta(0) + A_{0\theta}t \\
    & + A_{1\theta}\sin{(\omega_1t)} + A_{2\theta}\sin{(\omega_2t)} \\
    & + A_{3\theta}\sin{(2\omega_1t)} + A_{4\theta}\sin{(2\omega_2t)} \\
    & + A_{5\theta}\sin{\big[(\omega_1-\omega_2)t\big]} +A_{6\theta}\sin{\big[(\omega_1+\omega_2)t\big]}~~~,
\end{split}
\end{equation}

\noindent
where the coefficients $A_{j \theta}$ (given in Appendix \ref{app:evalthetaphi}) depend on the physical parameters and on $\beta$. The analytical formula of $\theta(t;\beta)$ can be used to directly evaluate the orbital period and hence, its variation over time $T(t)$. The latter is in good agreement with the numerical computations presented in \citep{Meyer2021}. The angle $\phi_1$ is given by

\begin{equation}
    \phi_1(t;\beta) = \phi_1(0) + \frac{p_{\phi_1}}{I_{1z}}t - \theta(t;\beta)~~~.
\end{equation}

The left panels in Figures \ref{fig4:set1b1} to \ref{fig7:set2b3} compare the solution (\ref{LinSol}) with the numerical one for two different values of $\beta$ and two parameter sets, as indicated in each Figure's caption. One can deduce that for $\beta = 1$, the solutions stay close to each other for several orbital periods (even months), but for $\beta = 3$, the solutions start deviating after a few periods. In fact, a gradual dephasing of the analytical from the numerical curves is present in all four cases, but eye-evident only in Figures \ref{fig6:set2b1} and \ref{fig7:set2b3}. Consequently, while the essential features of the motion and its basic parameters (amplitudes and frequencies) are determined relatively accurately by our linear model, they cannot be used always to express the long-term behavior of the perturbed system. For this purpose, a higher-order analytical theory is needed. In the following Section we present a more accurate model of the perturbed motion, using canonical perturbation theory.

\section{Nonlinear canonical perturbation theory}\label{sec:canonical}

\subsection{Hamiltonian preparation}\label{subsec:hamprep}

\noindent
In the previous Section, we discussed linear approximations to the post-deflection state computed via the equations of motion linearized around the new state $(r_{eq,new}, p_{\theta_{imp}})$. Nonlinear corrections to the above thereby can be obtained by considering terms of order higher than linear in the equations of motion, or higher than quadratic in the expansion of the Hamiltonian in powers of the quantities $(\delta r, \delta {\phi_2}, \delta p_r, \delta p_{\phi_2})$. In the present Section, we examine how to produce such nonlinear theories using the Lie series method of canonical perturbation theory. However, for algorithmic reasons, and also due to its necessity in the normal form theory developed for the full 3D triaxial model in a forthcoming paper, it turns convenient to first slightly modify the point of departure of such theories, i.e. the initial point around which the series expansions are to be made. To this end, consider a decomposition of the Hamiltonian
(\ref{HJ2ell}) as

\begin{equation}
    H = H_0 + H_1~~~,
\end{equation}

\noindent
where

\begin{equation}\label{H0}
    H_0 = \frac{p_r^2}{2m} +
          \frac{p_{\phi_1}^2}{2I_{1z}} +
          \frac{p_{\phi_2}^2}{2I_{2z}} +
          \frac{ (p_{\theta} - p_{\phi_1} - p_{\phi_2})^2 }{2mr^2} -
          \frac{GM_1M_2}{r}
\end{equation}

\begin{equation}
    H_1 = \frac{G M_1 ( I_{2x} + I_{2y} - 2I_{2z}) + 2GM_2(I_s - I_{1z}) }{4 r^3} +
                \frac{3GM_1 (I_{2x} - I_{2y}) \cos{(2\phi_2)} }{4 r^3}~~~.
\end{equation}

\noindent
The Hamiltonian $H_0$ describes the motion of two perfect spheres of masses $M_1,M_2$ and moments of inertia $I_{1z},I_{2z}$. The corresponding synchronous (Keplerian) equilibrium at separation $r = r^\ast$ of the binary is given by:

\begin{equation}\label{TwoSpheresEq}
\begin{gathered}
    r^\ast = r(0) \\
    \phi_2^\ast = 0 \\  
    p_r^\ast = 0 \\
    p_{\theta}^\ast = p_{\phi_1}^\ast + p_{\phi_2}^\ast + mr^{\ast 2}\nu_\theta^{\ast}  \\
    p_{\phi_1}^\ast = \nu_1 I_{1z} \\
    p_{\phi_2}^\ast = \nu_{\theta}^{\ast} I_{2z}~~~,
\end{gathered} 
\end{equation}

\noindent
with $\nu_1 = \dot{\theta}_{1z}(0)$ and $\nu_{\theta}^{\ast} = \sqrt{G(M_1+M_2)/r^{\ast 3}}$. Consider now an expansion of the Hamiltonian (\ref{H0}), up to a chosen truncation order $N + 2$, $N \geq 1$ in the variables $(\delta r, \delta {\phi_2}, \delta p_r, \delta p_{\phi_2})$ defined by the (canonical) transformation

\begin{equation}\label{CanDisturb}
\begin{gathered}
    r = r^\ast + \epsilon \cdot \delta r \\
    \phi_2 = \epsilon \cdot \delta \phi_2 \\
    p_r = \epsilon \cdot \delta p_r \\
    p_\theta = p_\theta^\ast + \epsilon^2 \cdot \delta p_\theta \\
    p_{\phi_1} = p_{\phi_1}^\ast + \epsilon^2 \cdot \delta p_{\phi_1} \\
    p_{\phi_2} = p_{\phi_2}^\ast + \epsilon \cdot \delta p_{\phi_2}~~~,
\end{gathered} 
\end{equation}

\noindent
where $\epsilon = 1$ is a `book-keeping' symbol. Following the `book-keeping' method (see \cite{Efthymiopoulos2012}), in all series expansions defined below we keep $\epsilon$ as a symbol, whose various powers collect terms representing a similar order of smallness of the problem. Since $(r^\ast, \phi_2^\ast = 0, p_r^\ast = 0, p_{\phi_2}^\ast; p_{\theta}^\ast, p_{\phi_1}^\ast)$ is an equilibrium of the 4x4 reduced set of Hamilton's equations, introducing (\ref{CanDisturb}) into the Hamiltonian $H_0$ and expanding in powers of $\epsilon$, leads to:

\begin{equation}\label{Heps}
\begin{split}
    H_0 & = \frac{GI_{2z}(M_1 + M_2) - GM_1M_2r^{\ast 2} + \nu_1^2I_{1z}r^{\ast 3}}{2r^{\ast 3}} \\
    & + \epsilon^2\bigg( \Pi_2(\delta r, \delta \phi_2, \delta p_r, \delta p_{\phi_2}) +
          \nu_{\theta}^{\ast}\delta p_\theta + 
          (\nu_1 - \nu_{\theta}^{\ast})\delta p_{\phi_1} + O(\epsilon) \bigg)~~~,
\end{split}
\end{equation}

\noindent
where $\Pi_2$ is a polynomial of second degree in its arguments. We find

\begin{equation}
    \Pi_2 = \frac{GM_1M_2}{2r^{\ast 3}}\delta r^2 +
          \frac{1}{2m} \delta p_r^2 +
          \bigg( \frac{1}{2I_{2z}} + \frac{1}{2mr^{\ast 2}} \bigg)\delta p_{\phi_2}^2 +
          \frac{2\nu_{\theta}^{\ast}}{r^\ast} \delta r \delta p_{\phi_2}~~~.
\end{equation}

\noindent
Since $\epsilon = 1$, we can lower at no cost all the book-keeping powers in the above expression by two, thus re-expressing $H_0$ as

\begin{equation}
\begin{split}
  H_0 & = \frac{GI_{2z}(M_1 + M_2) - GM_1M_2r^{\ast 2} + \nu_1^2I_{1z}r^{\ast 3}}{2r^{\ast 3}} \\ 
      &   + \frac{GM_1M_2}{2r^{\ast 3}}\delta r^2 +
          \frac{1}{2m} \delta p_r^2 +
          \bigg( \frac{1}{2I_{2z}} + \frac{1}{2mr^{\ast 2}} \bigg)\delta p_{\phi_2}^2 \\
          & + \frac{2\nu_{\theta}^{\ast}}{r^\ast} \delta r \delta p_{\phi_2} +
          \nu_{\theta}^{\ast}\delta p_{\theta} +
          (\nu_1 - \nu_{\theta}^{\ast})\delta p_{\phi_1} \\ 
      &   +\sum_{j=1}^{\infty} \epsilon^j H_{0, j+2}(\delta r, \delta \phi_2, \delta p_r, \delta p_{\phi_2};
             \delta p_\theta, \delta p_{\phi_1})~~~.
\end{split}
\end{equation}

\noindent
Since $\delta p_\theta, \delta p_{\phi_1}$ are integrals of motion, the constant terms $\nu_{\theta}^{\ast}\delta p_\theta + (\nu_1 - \nu_{\theta}^{\ast})\delta p_{\phi_1}$ can be omitted in the study of the reduced 4x4 dynamics in the variables $(\delta r, \delta \phi_2, \delta p_r, \delta p_{\phi_2})$. Keeping them emphasises, however, the fundamental difference between the degrees of freedom $(\theta, \delta p_\theta)$ and $(\phi_1, \delta p_{\phi_1})$, on one hand, and $(\delta r, \delta p_r)$, $(\delta \phi_2, \delta p_{\phi_2})$ on the other. Namely, the system is a rotator in the former, with frequencies $\nu_{\theta}$, $\nu_1 - \nu_{\theta}^{\ast}$, while it is an oscillator in the latter, with frequencies to be recovered by a diagonalization $+$ normalization procedure as described in the sequel. Also, keeping the term $(\nu_1 - \nu_{\theta}^{\ast})\delta p_{\phi_1}$ becomes essential once we relax the condition of axisymmetry of the primary, since then, the Hamiltonian becomes dependent on the angle $\phi_1$, and the above term properly enters the kernel of the normalization of the Hamiltonian (the term $\delta p_\theta$ still remains integral of motion in the full triaxial 3D case).

Returning, to the Hamiltonian term $H_1$, substituting (\ref{CanDisturb}) and expanding in powers of $\epsilon$, leads to:

\begin{equation}
\begin{split}
H_1 =& -\frac{G\big( M_2(I_{1z} - I_s) + M_1(-2I_{2x} + I_{2y} + I_{2z})\big)}{2r^{\ast 3}} \\
    & + \epsilon \frac{3G\big(M_2(I_{1z}-I_s) + M_1(-2I_{2x}+I_{2y}+I_{2z})\big)}{2r^{\ast 4}}\delta r \\
    & + \epsilon^2 \bigg[-\frac{3G\big(M_2 (I_{1z} - I_s) + M_1(-2I_{2x} + I_{2y} + I_{2z})\big)}{r^{\ast 5}} \delta r^2 +
        \frac{3GM_1 (I_{2y} - I_{2x})}{2r^{\ast 3}}\delta \phi_2^2 \bigg] \\
    & + \sum_{j=3}^{\infty} \epsilon^j H_{1,j}(\delta r, \delta \phi_2, \delta p_r,
                                               \delta p_{\phi_2}; \delta p_\theta, \delta p_{\phi_1})~~~.
\end{split}
\end{equation}

\noindent
Since $\epsilon = 1$, we lower again by two the book-keeping powers in the terms of $H_{1,j}$, but this time only for $j \geq 2$.  This leads to the book-kept Hamiltonian $H_1$

\begin{equation}
\begin{split}
H_1 =& -\frac{G\big( M_2(I_{1z} - I_s) + M_1(-2I_{2x} + I_{2y} + I_{2z})\big)}{2r^{\ast 3}} \\
    & -\frac{3G\big(M_2 (I_{1z} - I_s) + M_1(-2I_{2x} + I_{2y} + I_{2z})\big)}{r^{\ast 5}} \delta r^2 \\
    & + \frac{3GM_1 (I_{2y} - I_{2x})}{2r^{\ast 3}}\delta \phi_2^2 \\
    & + \epsilon \frac{3G\big(M_2(I_{1z}-I_s) + M_1(-2I_{2x}+I_{2y}+I_{2z})\big)}{2r^{\ast 4}}\delta r \\
    & + \sum_{j=1}^{\infty} \epsilon^j H_{1,j+2}(\delta r, \delta \phi_2, \delta p_r,
                                               \delta p_{\phi_2}; \delta p_\theta, \delta p_{\phi_1})~~~.
\end{split}
\end{equation}

\noindent
The final book-kept Hamiltonian has the form

\begin{equation}\label{Hexp}
    H = H_0 + H_1 =  Z_0 + \sum_{j=1}^{\infty} \epsilon^j h_j~~~,
\end{equation}

\noindent
where $Z_0$ and $h_j$ are obtained by adding the corresponding terms of $H_0$ and $H_1$ of the same book-keeping powers of $\epsilon$. Explicit expressions of the terms $Z_0$, $h_1$ and $h_2$ are given in the Appendix \ref{app:z0h1h2}.

The next step is to change our set of variables, from generalized positions and momenta $(\delta r, \delta {\phi_2}, \delta p_r, \delta p_{\phi_2})$ to Birkhoff complex canonical variables $(Q_1,Q_2,P_1,P_2)$, derived by diagonalizing the quadratic part of the Hamiltonian (\ref{Hexp}), i.e. the term $Z_0$. The relationship between $(\delta r, \delta {\phi_2}, \delta p_r, \delta p_{\phi_2})$ and $(Q_1,Q_2, P_1,P_2)$ is a linear canonical transformation

\begin{equation}\label{delta2QP}
\begin{bmatrix}
\delta r \\
\delta \phi_2 \\
\delta p_r \\
\delta p_{\phi_2}
\end{bmatrix} =
\begin{bmatrix}
m_{11} & m_{12} & m_{13} & m_{14} \\
m_{21} & m_{22} & m_{23} & m_{24} \\
m_{31} & m_{32} & m_{33} & m_{34} \\
m_{41} & m_{42} & m_{43} & m_{44}
\end{bmatrix}
\begin{bmatrix}
Q_1 \\
Q_2 \\
P_1 \\
P_2
\end{bmatrix}~~~,
\end{equation}

\noindent
where the elements $m_{ij}$ depend on the physical parameters of the system, as well as on the constant $r^{\ast}$. Detailed derivation of $m_{ij}$ is given in the Appendix \ref{app:QPtrans}. Substitution of (\ref{delta2QP}) into (\ref{Hexp}) leads to

\begin{equation}\label{HQP}
    H(Q_1,Q_2,P_1,P_2) = Z_0 + \sum_{j=1}^{N} \epsilon^j h_j(Q_1,Q_2,P_1,P_2)~~~,
\end{equation}

\noindent
where

\begin{equation}\label{Z0}
    Z_0 = C_{Z_0} + \nu_{\theta}^{\ast} \delta p_\theta + (\nu_1 - \nu_\theta^{\ast}) \delta p_{\phi_1} -
    i\omega_{1k} Q_1P_1 - i\omega_{2k} Q_2P_2~~~,
\end{equation}

\noindent
with

\begin{equation}\label{CZ0}
    C_{Z_0} = \frac{1}{2}\bigg( \nu_1^2I_{1z} + \frac{G\big( M_1(2I_{2x}-I_{2y} - M_2r^{\ast 2}) + M_2(-I_{1z}+I_{2z}+I_s) \big)}{r^{\ast 3}} \bigg)~~~,
\end{equation}

\noindent
and $\omega_{1k}$, $\omega_{2k}$ being the kernel frequencies (see expressions in the Appendix \ref{app:QPtrans}).

\subsection{Hamiltonian normal form and solution}\label{subsec:normalform}

\noindent
In order to extract analytical solutions from the book-kept Hamiltonian (\ref{HQP}), we apply a Birkhoff canonical normalization process using the method of Lie series \citep{Hori1966,Deprit1969} up to $N$ normalization steps. Detailed description of the process, in accordance with \cite{Efthymiopoulos2012} is presented in Appendix \ref{app:EvalNormalForm}. After performing $N$ steps of the normalization algorithm upon the Hamiltonian (\ref{HQP}), we obtain the normal form Hamiltonian

\begin{equation}\label{ExpandedZPol}
Z^{(N)} = \sum_{j_1=0}^{N_1 \leq N}\sum_{j_2=0}^{N_2 \leq N}\sum_{j_3=0}^{N_3 \leq N}\sum_{j_4=0}^{N_4 \leq N}C_{j_1j_2j_3j_4}^{(N)}(Q_1^{(N)})^{j_1}(P_1^{(N)})^{j_2}(Q_2^{(N)})^{j_3}(P_2^{(N)})^{j_4}~~~.
\end{equation}

\noindent
In equation (\ref{ExpandedZPol}), the constant coefficients $C_{j_1j_2j_3j_4}^{(N)}$ encapsulate the integrals of motion $\delta p_\theta$, $\delta p_{\phi_1}$, i.e. $C_{j_1j_2j_3j_4}^{(N)} = C_{j_1j_2j_3j_4}^{(N)}(\delta p_\theta, \delta p_{\phi_1})$, as well as the physical parameters. The variables $Q_j^{(N)}$ and $P_j^{(N)}$ appear only in the form of the products $Q_j^{(N)}P_j^{(N)}$, $j = 1,2$. The real quantities $I_j = iQ_j^{(N)}P_j^{(N)}$ are then integrals of motion under the flow of $Z^{(N)}$,  since

\begin{equation}
   \{ I_j \hspace{0.1cm} , \hspace{0.1cm} Z^{(N)} \} =
   \{ iQ_j^{(N)}P_j^{(N)} \hspace{0.1cm} , \hspace{0.1cm} Z^{(N)} \} = 0 , \hspace{0.15cm} j = 1,2~~~.
\end{equation}

\noindent
Due to the canonical nature of transformations, the preservation of the form of Hamilton's equations is achieved, thereby yielding

\begin{equation}\label{Hs4EOM}
\begin{gathered}
  \dot{Q}_1^{(N)} =  \frac{\partial Z^{(N)}}{\partial P_1^{(N)}} =  i\omega_{1c}Q_1^{(N)} \\
  \dot{Q}_2^{(N)} =  \frac{\partial Z^{(N)}}{\partial P_2^{(N)}} =  i\omega_{2c}Q_2^{(N)} \\
  \dot{P}_1^{(N)} = -\frac{\partial Z^{(N)}}{\partial Q_1^{(N)}} = -i\omega_{1c}P_1^{(N)} \\
  \dot{P}_2^{(N)} = -\frac{\partial Z^{(N)}}{\partial Q_2^{(N)}} = -i\omega_{2c}P_2^{(N)} \\
  \dot{\theta}^{(N)} =  \frac{\partial Z^{(N)}}{\partial \delta p_\theta} = \omega_{\theta c}   \\
  \dot{\phi}_1^{(N)} =  \frac{\partial Z^{(N)}}{\partial \delta p_{\phi_1}} = \omega_{\phi_1 c} ~~~, \\
\end{gathered} 
\end{equation}

\noindent
where the quantities $\omega_{1c}$, $\omega_{2c}$, $\omega_{\theta c}$ and $\omega_{\phi_1 c}$ contain only products of the form $Q_j^{(N)}P_j^{(N)}$, i.e. $\omega_{1c} = \omega_{1c}(I_1,I_2)$, $\omega_{2c} = \omega_{2c}(I_1,I_2)$, $\omega_{\theta c} = \omega_{\theta c}(I_1,I_2)$ and $\omega_{\phi_1 c} = \omega_{\phi_1 c}(I_1,I_2)$. It follows that equations (\ref{Hs4EOM}) can be directly integrated, giving the solutions

\begin{equation}\label{FinSolQP4}
\begin{gathered}
  Q_1^{(N)}(t) = Q_1^{(N)}(0)e^{ i\omega_{1c}t } \\
  Q_2^{(N)}(t) = Q_2^{(N)}(0)e^{ i\omega_{2c}t } \\
  P_1^{(N)}(t) = P_1^{(N)}(0)e^{-i\omega_{1c}t } \\
  P_2^{(N)}(t) = P_2^{(N)}(0)e^{-i\omega_{2c}t } \\
  \theta^{(N)}(t) = \theta^{(N)}(0) + \omega_{\theta c}t \\
  \phi_1^{(N)}(t) = \phi_1^{(N)}(0) + \omega_{\phi_1 c}t ~~~,
\end{gathered} 
\end{equation}

\noindent
where $\big(Q_1^{(N)}(0),Q_2^{(N)}(0),P_1^{(N)}(0),P_2^{(N)}(0)\big)$ are the initial conditions of any trajectory transformed to the Birkhoff variables, and subjected to $N$  transformations with the Lie series. In Hamilton's equations, we include the variables $\big( \theta^{(N)}, \phi_1^{(N)} \big)$, which, on the one hand are subjected to the Lie series transformations, but on the other hand do not affect the Hamiltonian normalization process since they are ignorable coordinates. Mapping the solution (\ref{FinSolQP4}) into the original variables $\big(\delta r(t), \delta \phi_2(t), \delta p_r(t), \delta p_{\phi_2}(t)\big)$ and $\big( \theta(t), \phi_1(t) \big)$, yields the final solution

\begin{equation}\label{BetaCanSol}
\begin{gathered}   
r_{can}(t;\beta) =
r^{\ast} + P_{r,0}(\beta) +
\sum_{j=1}^{\omega_{r,tot}} C_{r,j}\beta^{k_{r,j}}\cos{\big[(n_j\omega_{1c} + m_j\omega_{2c})t\big]} \\
\phi_{2can}(t;\beta) =
\sum_{j=1}^{\omega_{\phi_2,tot}} C_{\phi_2,j}\beta^{k_{\phi 2,j}}\sin{\big[(n_j\omega_{1c} + m_j\omega_{2c})t\big]}\\
p_{r_{can}}(t;\beta) =
\sum_{j=1}^{\omega_{p_r,tot}} C_{p_r,j}\beta^{k_{pr,j}}\sin{\big[(n_j\omega_{1c} + m_j\omega_{2c})t\big]} \\
p_{\phi_{2can}}(t;\beta) =
p_{\phi_2}^{\ast} + P_{p_{\phi_2},0}(\beta) +
\sum_{j=1}^{\omega_{p_{\phi_2},tot}}C_{p_{\phi_2},j}\beta^{k_{p\phi2,j}}\cos{\big[(n_j\omega_{1c}+m_j\omega_{2c})t\big]} \\
\theta_{can}(t;\beta) = \theta(0) + \omega_{\theta c}(\beta) t +
\sum_{j=1}^{\omega_{\theta,tot}} C_{\theta,j}\beta^{k_{\theta,j}}\sin{\big[(n_j\omega_{1c} + m_j\omega_{2c})t\big]} \\
\phi_{1can}(t;\beta) = \phi_1(0) + \frac{p_{\phi_1}}{I_{1z}}t - \theta_{can}(t;\beta)~~~.
\end{gathered}
\end{equation}

\noindent
In equation (\ref{BetaCanSol}), the frequencies $\omega_{1c}, \omega_{2c}, \omega_{\theta c}$ and $\omega_{\phi_1 c}$ are polynomial functions of $\beta$

\begin{equation}\label{nu12}
\begin{gathered}
\omega_{1c}(\beta) = \Omega_{10} + \Omega_{11}\beta + ... + \Omega_{1N}\beta^N + ... \\
\omega_{2c}(\beta) = \Omega_{20} + \Omega_{21}\beta + ... + \Omega_{2N}\beta^N + ... \\
\omega_{\theta c}(\beta)=\Omega_{\theta 0}+\Omega_{\theta 1}\beta+...+\Omega_{\theta N}\beta^N + ... \\
\omega_{\phi_1 c}(\beta)=\Omega_{\phi_1 0}+\Omega_{\phi_1 1}\beta+..+\Omega_{\phi_1 N}\beta^N + ...~~~,
\end{gathered}
\end{equation}

\noindent
with $\Omega_{1j},\Omega_{2j},\Omega_{\theta j},\Omega_{\phi_1 j}$ depending on the physical parameters. Also, the polynomials $P_{r,0}(\beta)$ and $P_{p_{\phi_2},0}(\beta)$ have the form

\begin{equation}
\begin{gathered} 
     P_{r,0}(\beta) = D_{r,0} + D_{r,1}\beta + ... + D_{r,N}\beta^N + ... \\
     P_{p_{\phi_2},0}(\beta) = D_{p_{\phi_2},0} + D_{p_{\phi_2},1}\beta + ... + D_{p_{\phi_2},N}\beta^N + ...
\end{gathered}
\end{equation}

\noindent
where $D_{r,j}$, $D_{p_{\phi_2},j}$ depend on the physical parameters as well. The numbers $n_j,m_j$ are integers that constitute combinations of $\omega_{1c}$ and $\omega_{2c}$. The counters $\omega_{r,tot}$, $\omega_{\phi_2,tot}$, $\omega_{p_r,tot}$, $\omega_{p_{\phi_2},tot}$ and $\omega_{\theta,tot}$ are the total number of combinations of $(n_j,m_j)$, correspondingly for the functions $r_{can}(t;\beta)$, $\phi_{2can}(t;\beta)$, $p_{r_{can}}(t;\beta)$, $p_{\phi_{2can}}(t;\beta)$ and $\theta_{can}(t;\beta)$. The amplitude coefficients $C_{r,j}$, $C_{\phi_2,j}$, $C_{p_r,j}$, $C_{p_{\phi_2},j}$ and $C_{\theta,j}$ depend on the physical parameters.  Comparing equations (\ref{LinSol}) and (\ref{BetaCanSol}), we can see that the canonical perturbation theory adds corrections to the functions $r(t;\beta)$, $\phi_2(t;\beta)$, $p_r(t;\beta)$ and $p_{\phi_2}(t;\beta)$ through higher order polynomials of $\beta$. Those corrections affect the amplitudes $(C_{r,j}\beta^{k_{r,j}}$, $C_{\phi_2,j}\beta^{k_{\phi 2,j}}$, $C_{p_r,j}\beta^{k_{pr,j}}$, $C_{p_{\phi_2},j}\beta^{k_{p\phi2,j}})$, the frequencies $(n_j\omega_{1c}(\beta) + m_j\omega_{2c}(\beta))$ and the equilibrium point $(r^\ast + P_{r,0}(\beta)$,  $p_{\phi_2}^{\ast} + P_{p_{\phi_2},0}(\beta))$ for which our initial guess is the Keplerian one. What's more, the canonical theory corrects the orbital angle $\theta$ (by which the orbital period can be measured), by changing the Keplerian orbital rate from $\nu_\theta$ to $\omega_{\theta c}(\beta)$ and by adding appropriate trigonometric variations.

\subsection{Numerical comparisons and precision tests}\label{subsec:comparisons}

\begin{figure}[!h]
  \centering
    \includegraphics[scale=0.5]{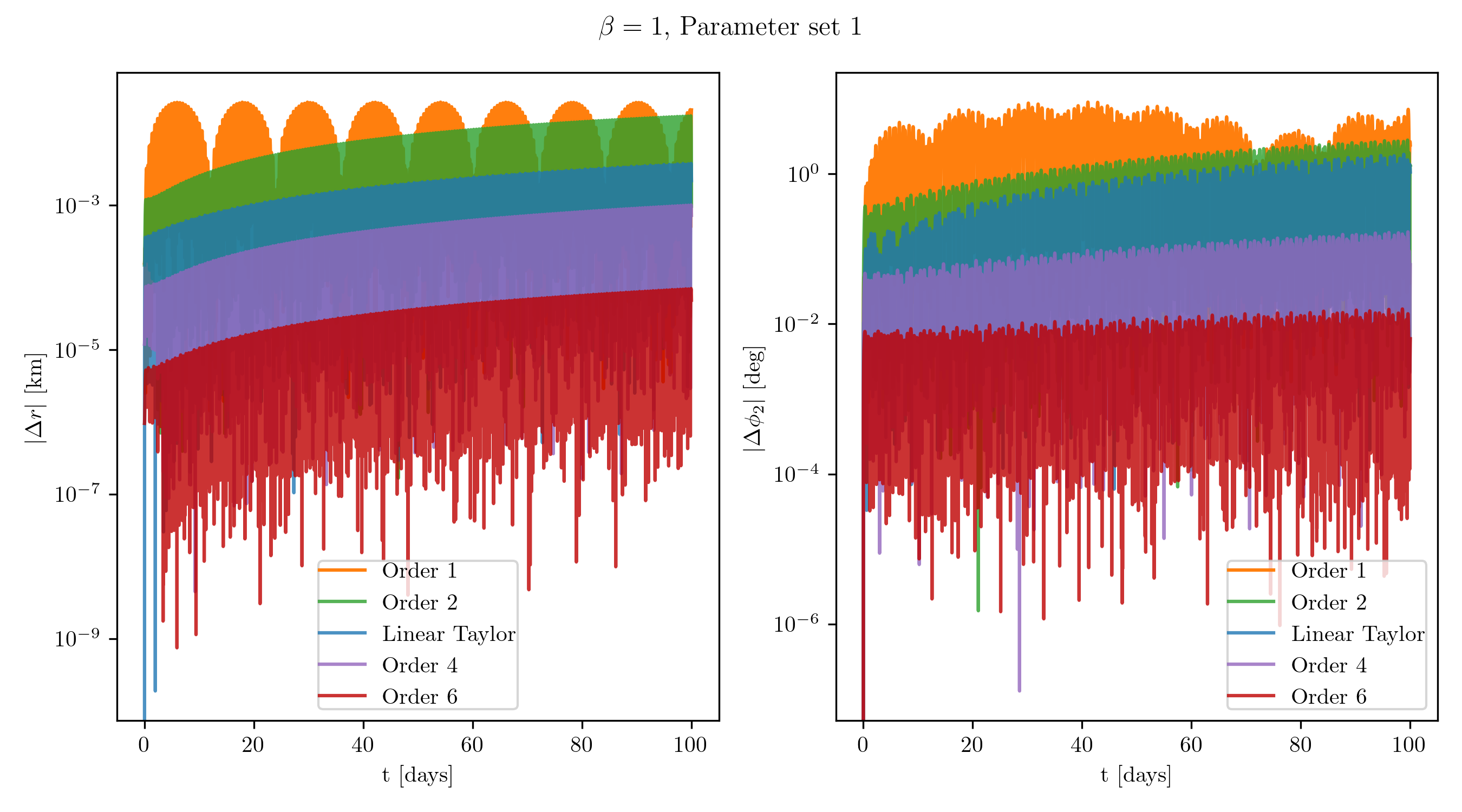}
    \includegraphics[scale=0.5]{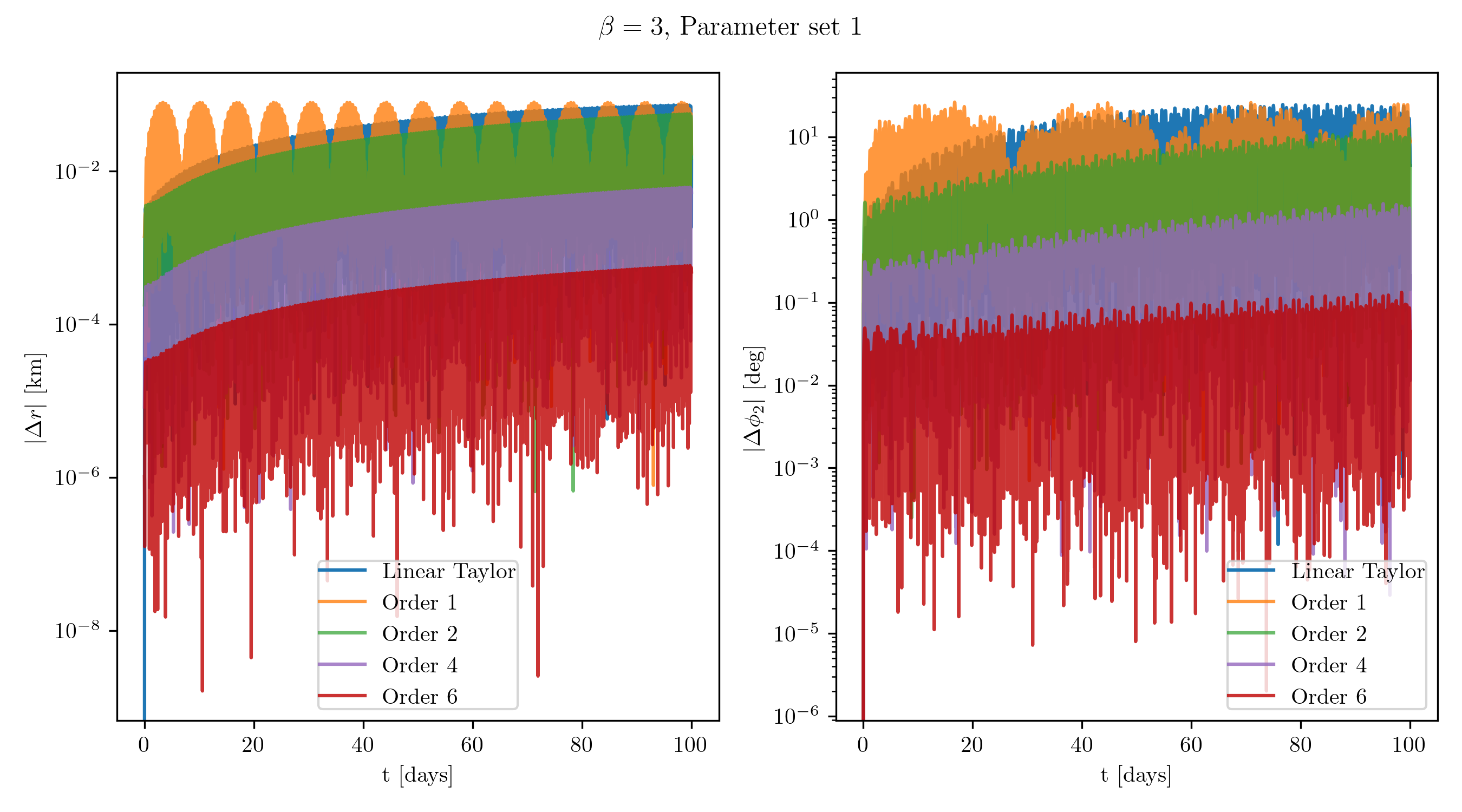}
    \caption{Absolute differences between the numerical and analytical solutions of the
    functions $r(t)$ and $\phi_2(t)$ for a total duration of 100 days. For the first set
    of physical parameters, the cases for $\beta = 1,3$ are shown.}
    \label{fig9:drphiset1}
\end{figure}

\begin{figure}[!h]
  \centering
    \includegraphics[scale=0.5]{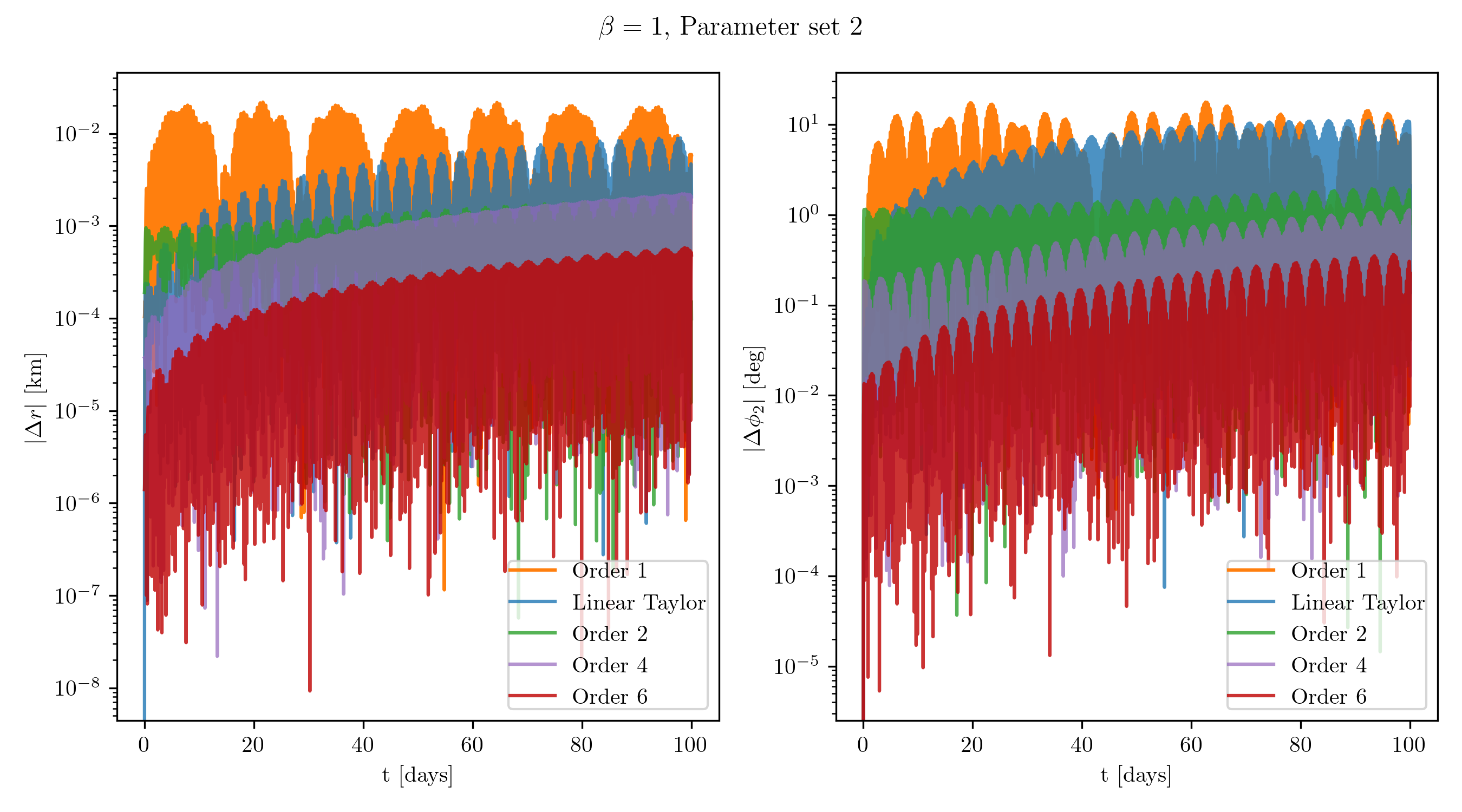}
    \includegraphics[scale=0.5]{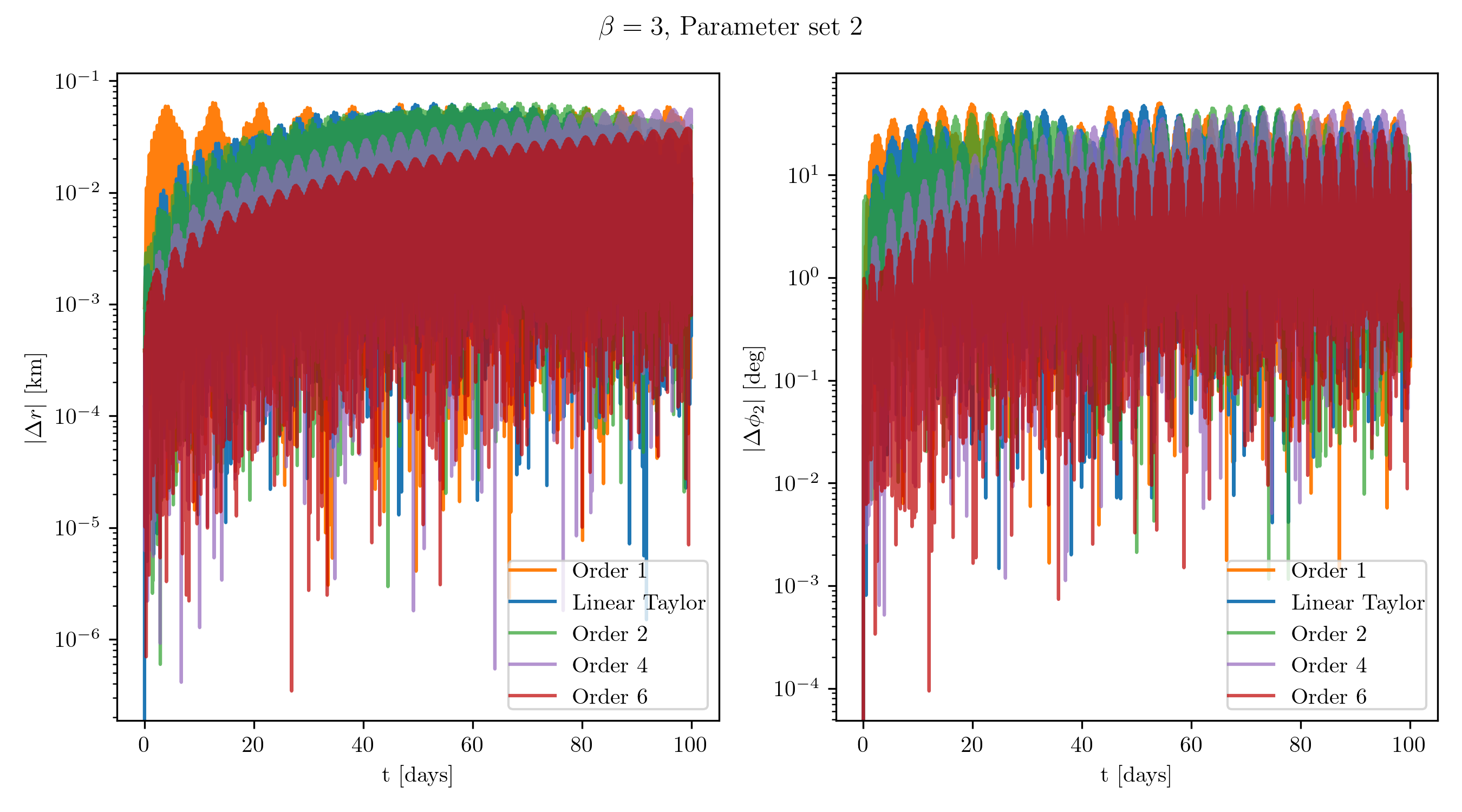}
    \caption{Same plots as in Figure \ref{fig9:drphiset1}, but for the second set of physical parameters.}
    \label{fig10:drphiset2}
\end{figure}

So far we have examined two types of theories by which we can obtain a semi-analytic representation of the solutions $\delta r(t),\delta\phi_2(t), \delta p_r(t),\delta p_{\phi_2}(t),\theta(t),\phi_1(t)$ for perturbed librational motions around the isochronous equilibrium state: i) the `Linear Taylor' theory of Section \ref{sec:linear}, and ii) the nonlinear theories of the present section. These theories can be diversified in two aspects. First, the linear theory of Section \ref{sec:linear} differs from the normal form theories of the present section in that the former is computed around the new, post-impact, equilibrium, of radius $r_{eq,new}$ (see Section \ref{sec:linear}), while the latter introduce corrections to a theoretical equilibrium of the Keplerian problem at $r=r^{\ast}$ (see Subsection \ref{subsec:hamprep}). This implies that the linear theory of the previous section can be better in accuracy than the low order normal form theories (e.g. $N=1,2$). However, normal form theories of higher truncation orders eventually take over as regards the accuracy. Below we examine the accuracy of the series solutions of orders $N=1,2,4,6$. A visual comparison, in particular of Figures \ref{fig5:set1b3}, \ref{fig6:set2b1} and \ref{fig7:set2b3}, shows that the nonlinear theories can fit the waveforms of the curves of the numerical solutions, better than the linear theory, in most cases already at truncation $N=2$. A more detailed comparison is shown in Figures \ref{fig9:drphiset1} and \ref{fig10:drphiset2} (errors in $r(t)$ on the left and errors in $\phi_2(t)$ on the right).

\noindent
As regards the `Linear-Taylor' and the $N=2$ theories, in general the latter performs better than the former, but with noticeable exceptions (e.g. in the curves $r(t)$, $\phi_2(t)$ in Figure \ref{fig9:drphiset1} for $\beta = 1$). We attribute this to the better definition of the equilibrium state for the linear theory, which (depending on the magnitude of the perturbation and the physical parameters) can only be superseded by orders higher than second in perturbation theory.

\begin{figure}[h!]
    \centering
	\includegraphics[scale=0.5]{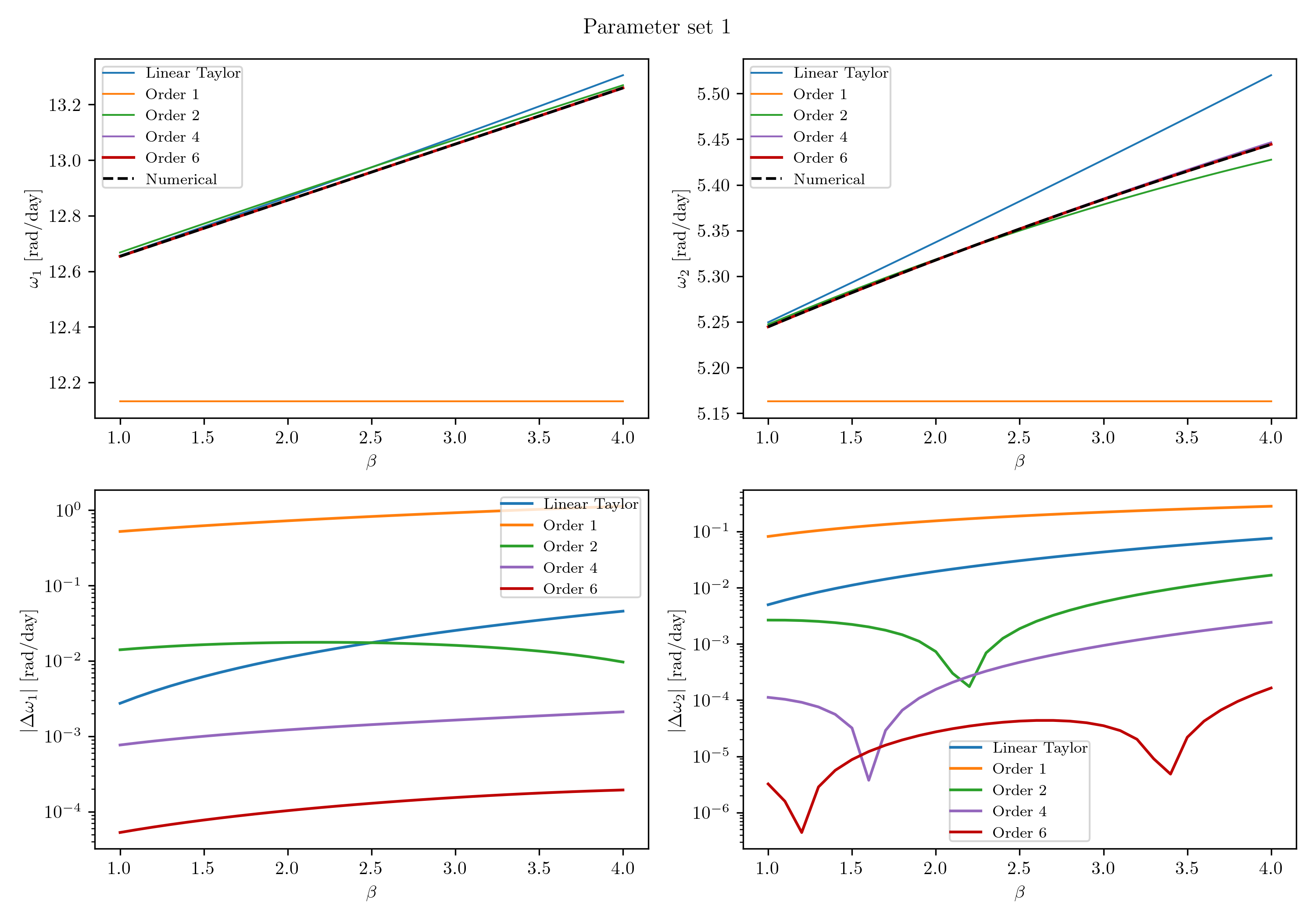}
    \includegraphics[scale=0.5]{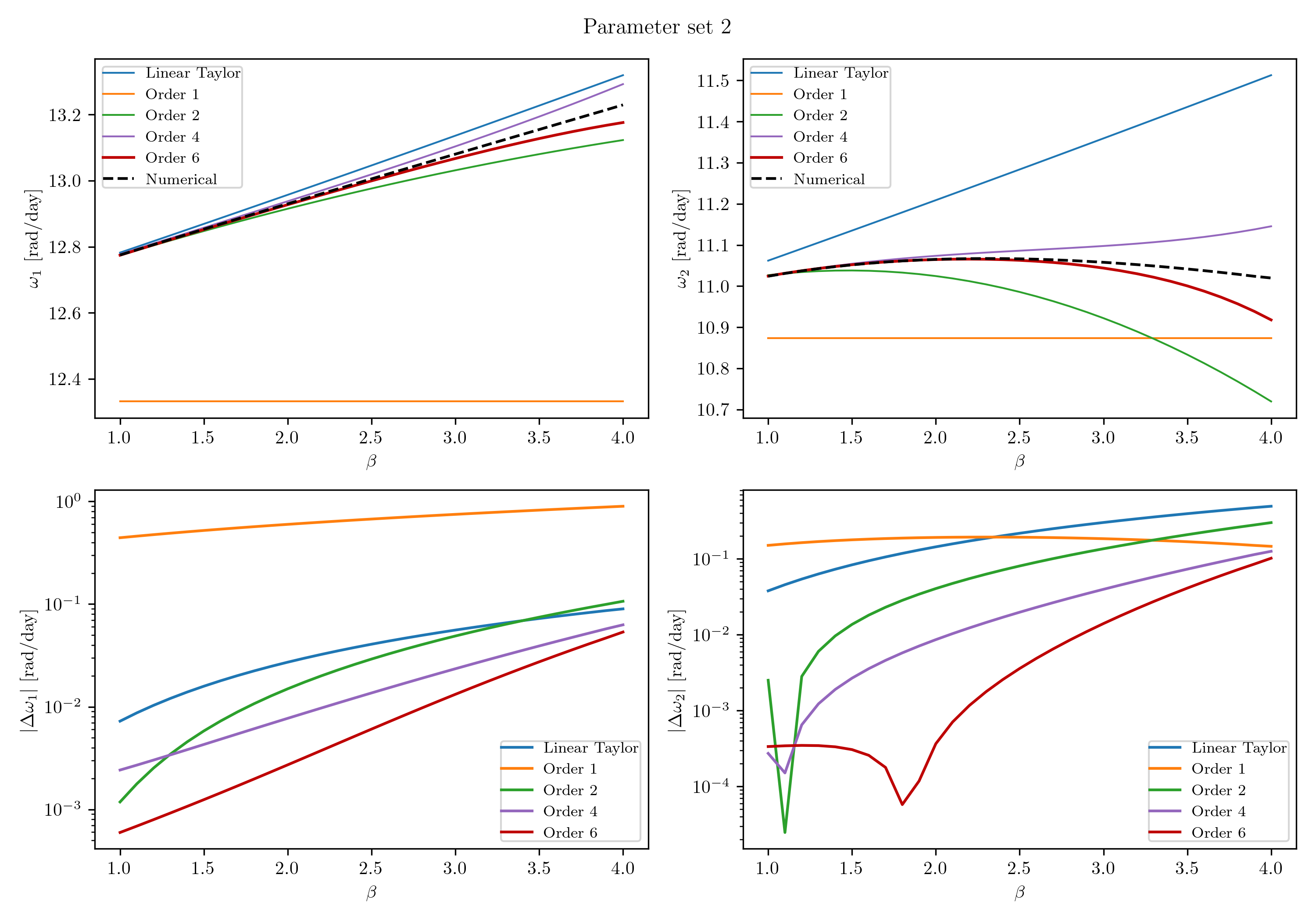}
    \caption{Comparison of the frequencies derived by NAFF $(\omega_{1,num}$, $\omega_{2,num})$ with
    those of analytical theories $(\omega_1$, $\omega_2$, $\omega_{1c}$, $\omega_{2c})$ for
    different values of $\beta$.}
    \label{fig11:omegas}
\end{figure}

It is clear that the systematic growth of errors in Figures \ref{fig9:drphiset1} and \ref{fig10:drphiset2} reflects mostly the error in the determination of the fundamental frequencies $\omega_{1,num}$ and $\omega_{2,num}$ through any of the examined theories. The latter can be observed in Figure \ref{fig11:omegas} which depicts the frequencies themselves as functions of the $\beta$ parameter. Apart from the black curve (numerical), all others refer to the frequency evaluation by each of the theories in consideration. The values that correspond to the black curves (numerical), are obtained by performing Laskar's NAFF \citep{Laskar1992} on the timeseries of Figures \ref{fig4:set1b1} to \ref{fig7:set2b3}. We now see more clearly the improvement of the precision with the order of the theory. Note, in particular, that for the examined range of $\beta$, we typically have small relative errors, implying that the corresponding analytical formulas can be used to fit $\beta$ from observed time series of length, even as high as of several months or years. Linear theory, instead, can be used for fitting $\beta$ from observed time series of a limited span of few days.

\begin{figure}[h!]
    \centering
	\includegraphics[scale=0.4]{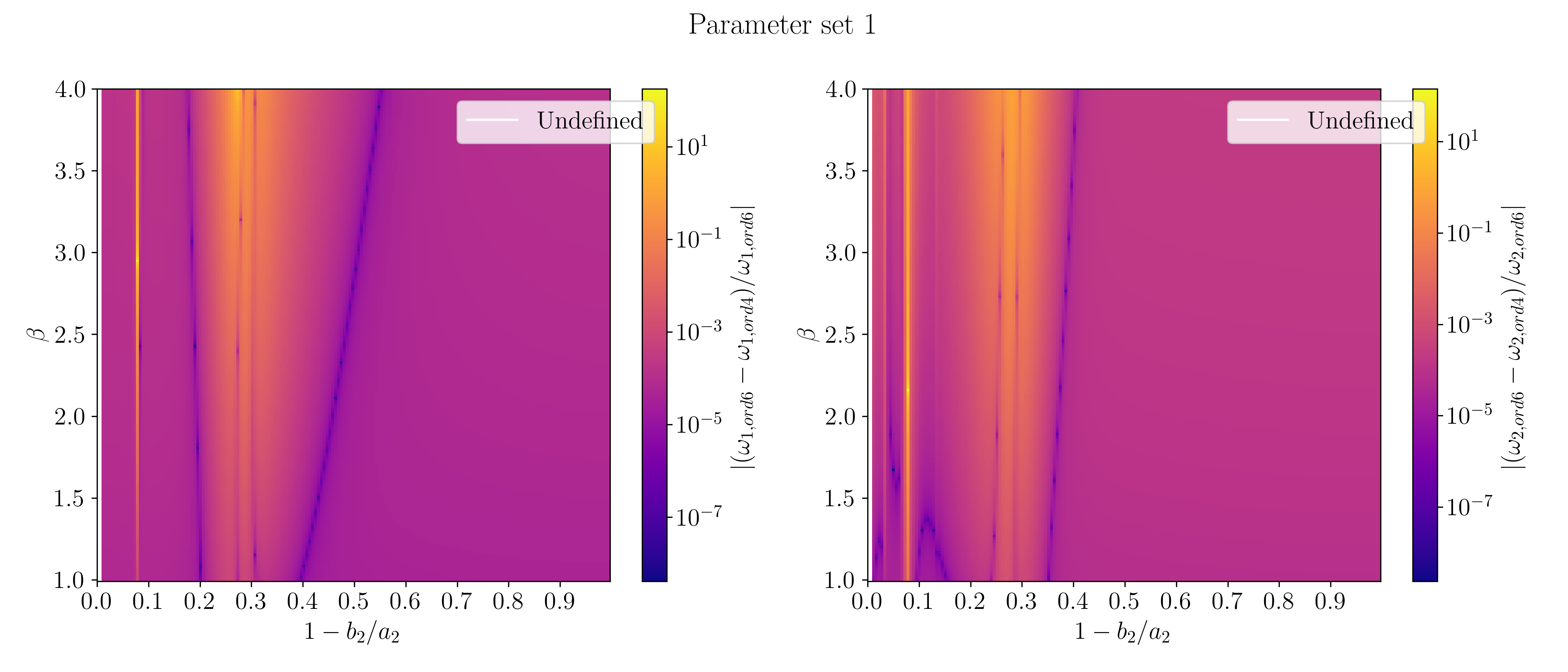}
    \includegraphics[scale=0.4]{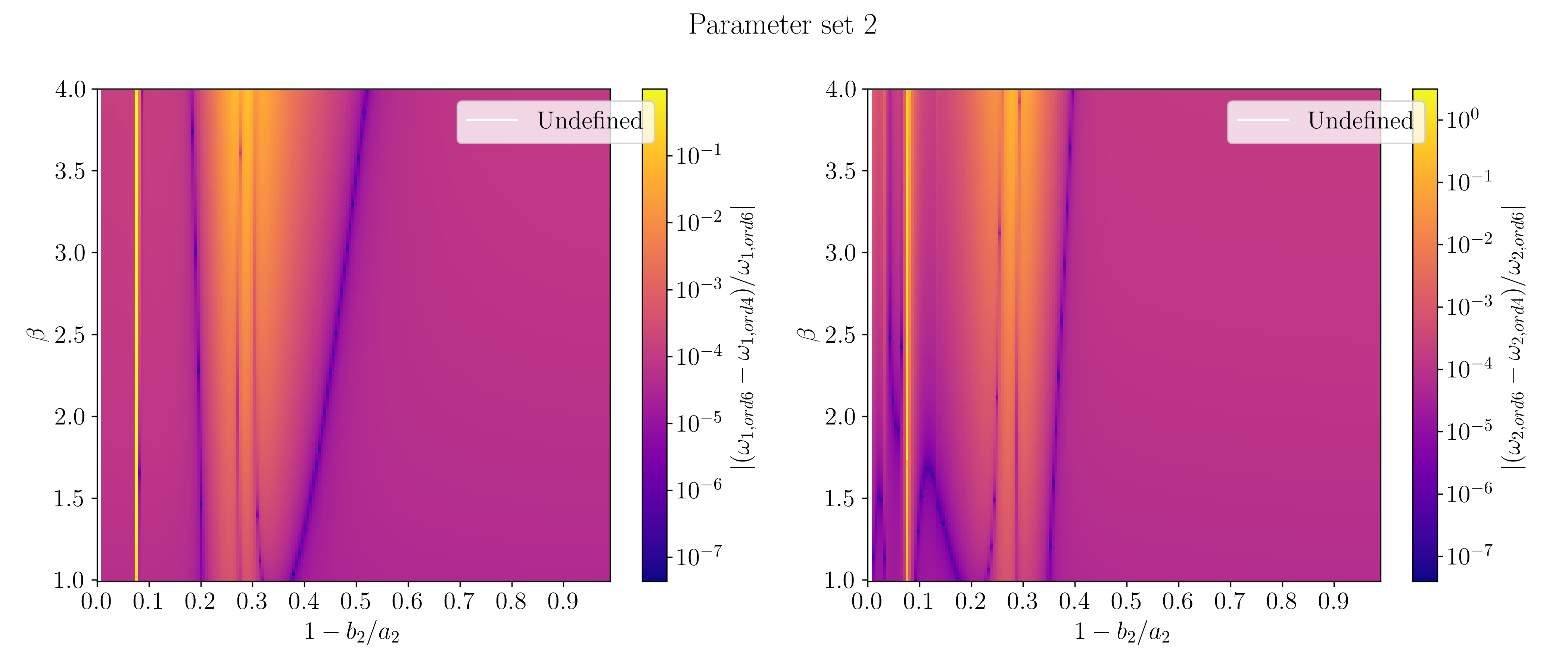}
    \caption{Relative errors in the analytical frequencies, as derived via the $4_{th}$ and the $6_{th}$ order theory.}
    \label{fig12:colormaps}
\end{figure}

\begin{figure}[h!]
    \centering
	\includegraphics[scale=0.31]{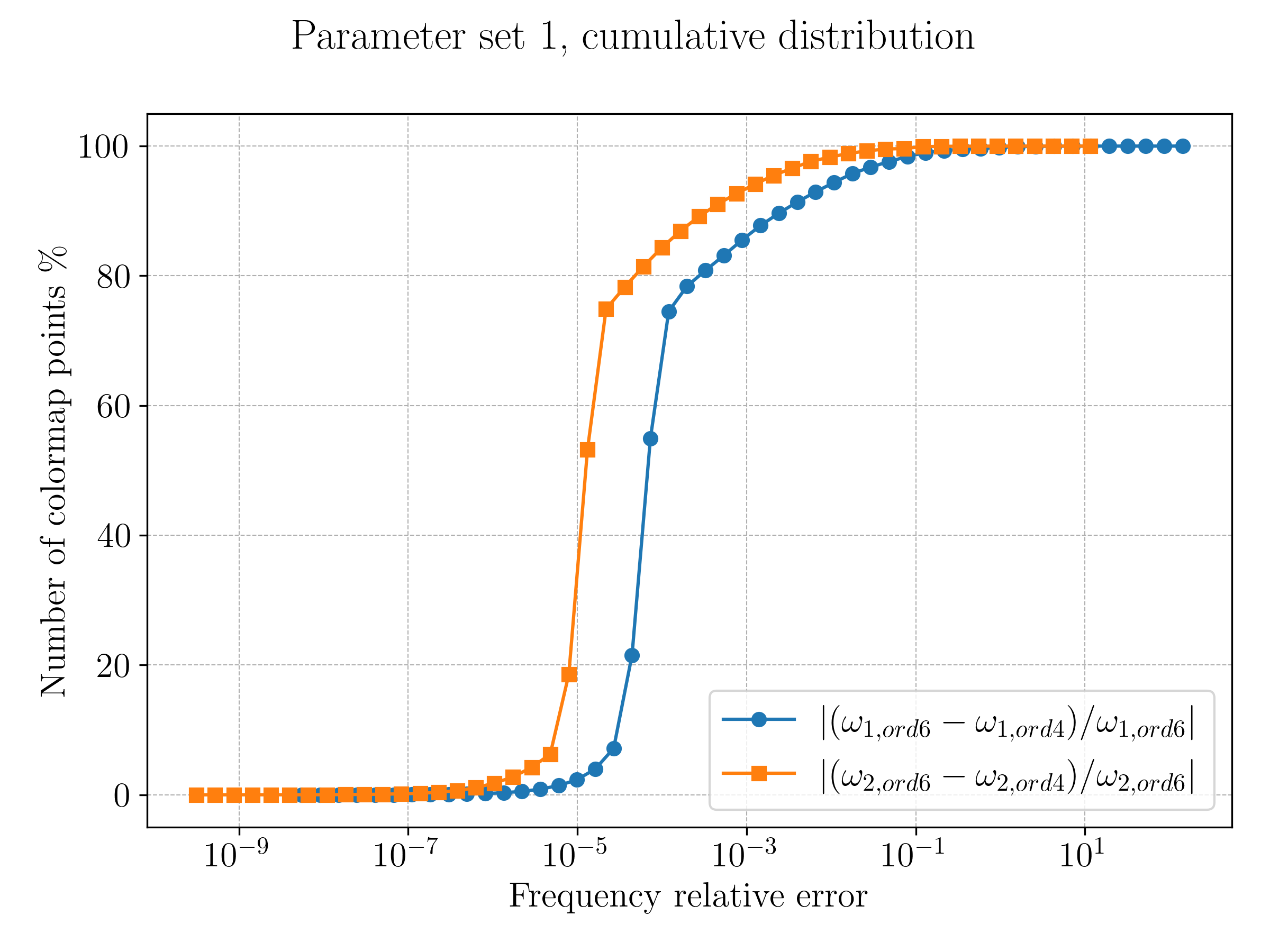}
    \includegraphics[scale=0.31]{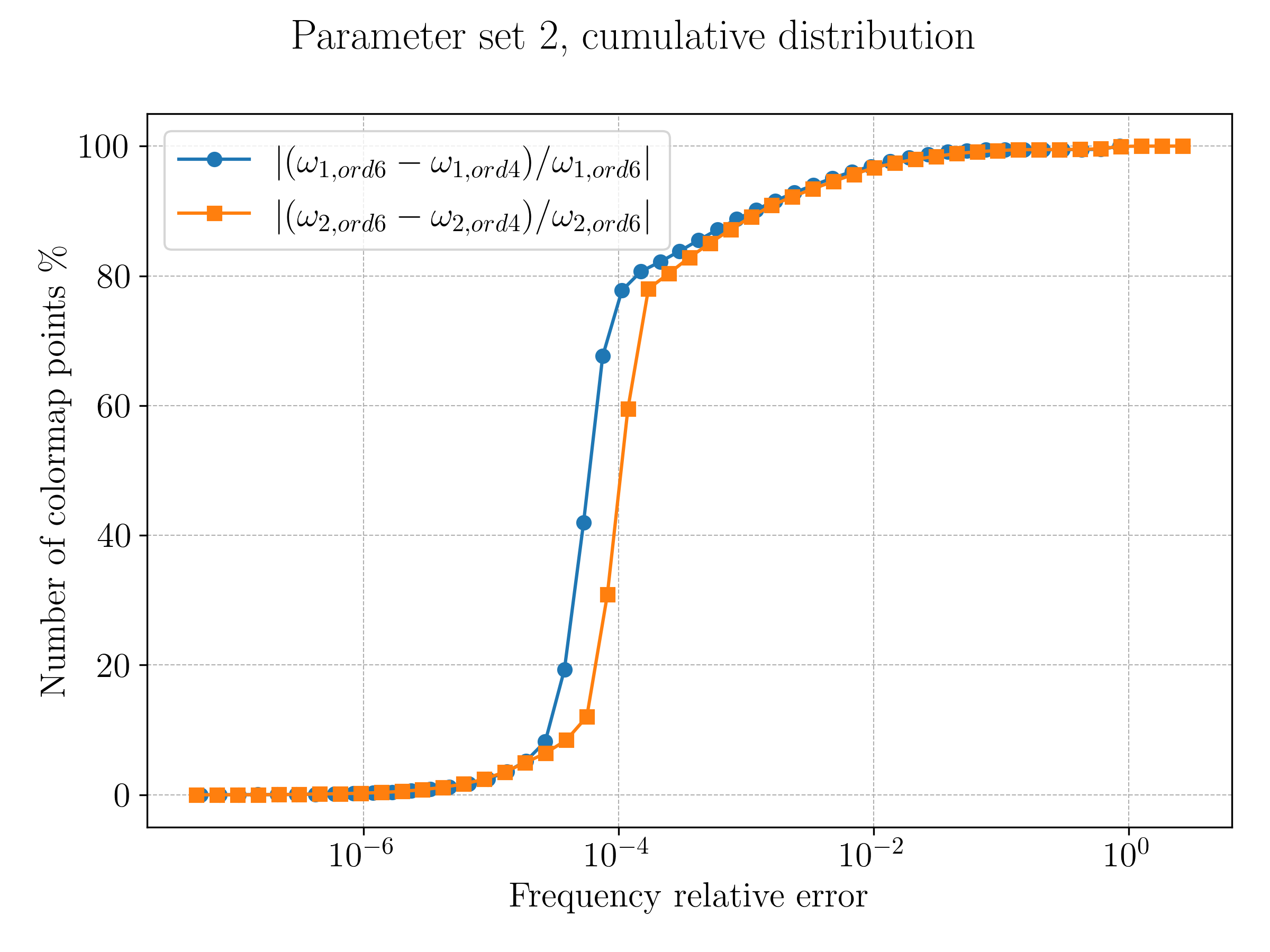}
    \caption{Cumulative distributions of the relative errors of the analytical frequencies that correspond to Figure \ref{fig12:colormaps}.}
    \label{fig13:cumdistros}
\end{figure}

An estimate of the convergence properties, as well as the error of the normal form approximation, can be obtained without comparison with the numerical simulations, by comparing several quantities, depending on the order of the normal form truncation, as shown in Figure \ref{fig12:colormaps}. This figure shows the relative differences in the prediction of the analytical frequencies $\omega_1$, $\omega_2$, as derived from the $4_{th}$ and $6_{th}$ order normal form theories, for the two parameter sets, plotted against $\beta$ and the secondary asteroid's asphericity $(1 - b_2/a_2)$. The figure serves to distinguish the regions where the theory correctly predicts (up to a predefined accuracy) the frequencies, from those where it does not. Although the check performed in Figure \ref{fig12:colormaps} can be repeated for a variety of physical parameters which enter into the problem, such as the masses and the moments, our focus on $\beta$ and the secondary's asphericity stems from their critical influence on the dynamics of the DART experiment. The yellow regions in the colormap indicate areas of high frequency errors. These regions correspond to resonances between the natural frequencies of the system, which in turn may lead to chaotic evolution (\cite{Agrusa2021}). The polynomial expressions (\ref{nu12}) diverge at these parameter regions, causing the predicted frequencies to become unbounded. As a result, the relative differences increase significantly, indicating the overall failure of the approximation based on the perturbation theory. Figure \ref{fig13:cumdistros} depicts the cumulative distributions of the percentage of the area (grid points) of the parameter space of Figure \ref{fig12:colormaps}, where the relative error of the method is smaller than the value indicated in the abscissa. By selecting an upper bound for the error on the $x-$axis, one may determine the proportion of colormap grid points, represented on the $y-$axis, that fall within this error threshold. As shown in these plots, for both our chosen sets of parameters, the normal form method produces a relative error of $10^{-5}$ or smaller for about $20\%$ of the grid points, while this percentage grows to about $80\%$ accepting a relative error smaller than $10^{-2}$.

\section{Conclusions}\label{sec:conclusions}

In the present paper we developed linear and nonlinear (canonical perturbation) theories able to describe by explicit formulas in time the evolution of the librational state of a binary deflected via a DART-like hit. We also gave examples of application of such theories for parameter values pertinent to the case of the 65803 Didymos system. The main steps and conclusions in our study are given below.

We assumed a second-moment multipole expansion of the mutual potential and introduced two basic assumptions: planar motion and axisymmetry of the primary. The first of these assumptions can be removed, at the cost of larger complexity in the computations, as will be discussed in a forthcoming work. The second assumption can also be removed by adding the degree of freedom $\phi_1$ to the system. However, we argued that this is not really necessary when the primary asteroid is a fast rotator, since, then, at formal level the assumption of axisymmetry is similar to first order averaging theory with respect to the fast angle $\phi_1$. On the other hand, the formulas introduced here are straightforward to generalize in the genuine case of a triaxial primary.

In all examined theories, the corresponding explicit formulas representing the time series of any post-impact observable quantity of the trajectories (i.e. depending on $r(t)$ and $\phi_2(t)$) were given in parametric form in the momentum transfer parameter $\beta$. This is crucial for the above-reported methods to be of utility in data fitting obtained by observations of the system after the impact. Of course, the obvious advantage of such an analytical approach is that an analytical formula can be introduced at nearly zero computational cost in the least-square fitting algorithm, and can lead even to direct formulas for the optimal estimator $\hat{\beta}$ of $\beta$, while, on the contrary, any fitting method based on numerical trajectories requires running the whole trajectory anew per trial value of $\beta$. We may also imagine several additional applications of the analytical formulas, i.e., in the modelling of Earth-observed light curves of the binary asteroid, the detection of non-uniform motions, variations in the orbital eccentricity, etc.

As would be expected, nonlinear normal form theories give better levels of precision as the order of the theory increases. On the other hand, truncation beyond a certain order, quickly becomes non-practical, as the number of series terms quickly proliferates. We found that our theories give small errors in the determination of the fundamental frequencies of the librational and epicyclic motion, which are the main factors affecting the systematic growth of errors in the curves $r(t)$ and $\phi_2(t)$, caused by systematic drifts in the phase of the oscillation. On the other hand, even linear theory can give small errors in the frequencies determination, provided that the determination of the initial synchronous state around which the theory is developed is precise. Finally, the errors grow in general with $\beta$, as larger $\beta$ implies in general larger post-impact libration amplitudes for the system, hence, a larger importance of the nonlinear corrections to the equations of motion\footnote{The symbolic codes used to produce the normal form series expansions of the present paper are available upon request to the authors.}.


\newpage

\backmatter

\bmhead{Acknowledgements}

M.G., I.G., G.V. and K.T acknowledge funding support from the European Union’s Horizon 2020 research and innovation program under grant agreement No. 581870377 (project NEO-MAPP). 

\bmhead{Conflicts of interest}

M.G., I.G. and G.V. have no conflicts of interest to declare that are relevant to the content of this article. C.E. and K.T. are Editorial Board Members of Celestial Mechanics and Dynamical Astronomy.

\begin{appendices}

\section{Nonzero elements of the Jacobian matrix $\boldsymbol{J}$ }\label{app:jnm}

\begin{equation}
    j_{13} = \frac{1}{m}
\end{equation}

\begin{equation}
    j_{21} = \frac{\sqrt{2G\left(3M_2(I_{1z} - I_s) +
             3M_1(-2 I_{2x} + I_{2y} + I_{2z}) + 2M_1M_2 r_{eq,new}^2\right)}}{\sqrt{m r_{eq,new}^7}}
\end{equation}

\begin{equation}
    j_{24} = \frac{1}{I_{2z}}+\frac{1}{m r_{eq,new}^2}
\end{equation}

\begin{equation}
    j_{31} = \frac{G \left(3 M_2 (I_{1z} - I_s) +
             3M_1 (-2I_{2x} + I_{2y} + I_{2z}) - 2M_1M_2 r_{eq,new}^2\right)}{2r_{eq,new}^5}
\end{equation}

\begin{equation}
    j_{42} = \frac{3GM_1(I_{2x} - I_{2y})}{r_{eq,new}^3}
\end{equation}


\section{Coefficients of the characteristic polynomial}\label{app:xi02}

\begin{equation}
\begin{split}
\xi_2 & = \frac{ 3G\big(M_2(I_s - I_{1z}) + M_1(I_{2y} - I_{2z})\big) }{2mr_{eq,new}^5} \\
      & + \frac{G \big( M_1(-3I_{2x} + 3I_{2y} + I_{2z}) + M_2I_{2z} \big) }{I_{2z}r_{eq,new}^3}
\end{split}
\end{equation}

\begin{equation}
\begin{split}
\xi_0 & = \frac{ -3G^2(I_{2x} - I_{2y})(M_1 + M_2)}{2I_{2z}M_1M_2^2 r_{eq,new}^{10}} \\
      &   \times \bigg[ -3M_1M_2 r_{eq,new}^2 \big(M_2 ( I_{1z} + 2I_{2z} - I_s ) \\
      &   + M_1(-2 I_{2x} + I_{2y} + 3I_{2z}) \big) \\
      &   - 15I_{2z}\big(M_1 + M_2\big)\big(M_2(I_{1z} - I_s) \\
      &   + M_1(-2I_{2x} + I_{2y} + I_{2z})\big) + 2M_1^2M_2^2r_{eq,new}^4 \bigg]
\end{split}
\end{equation}


\section{Amplitude coefficients of the linear solution}\label{app:lincoeff}

\begin{equation}
    A_{1r} =
    \frac{ \delta p_{\phi_2}(0)j_{13}j_{21} - \omega_2^2\delta r(0) - j_{13}j_{31}\delta r(0)}{\omega_1^2-\omega_2^2}
\end{equation}

\begin{equation}
    A_{2r} =
    \frac{-\delta p_{\phi_2}(0)j_{13}j_{21} + \omega_1^2\delta r(0) + j_{13}j_{31}\delta r(0)}{\omega_1^2-\omega_2^2}
\end{equation}

\section{DART-like disturbance at $t = 0$}\label{app:dartlike}

\begin{equation}
    \delta r(0) = r_{eq} - r_{eq,new}
\end{equation}

\begin{equation}
    \delta p_{\phi_2}(0) = [\dot{\theta}_{eq}(r_{eq}) - \dot{\theta}_{eq,new}(r_{eq,new})]I_{2z} 
\end{equation}

\begin{equation}
    \delta p_r(0) = \delta \phi_2(0) = 0 
\end{equation}


\section{Evaluation of the functions $\theta(t)$ and $\phi_1(t)$}\label{app:evalthetaphi}

\noindent
At any moment, the orbital displacement rate $\dot{\theta}$ is given by

\begin{equation}\label{EqTheta}
    \dot{\theta} = \frac{p_\theta - p_{\phi_1} - p_{\phi_2}(t)}{mr(t)^2} \Rightarrow
    \theta(t) = \int \frac{p_\theta - p_{\phi_1} - p_{\phi_2}(t)}{mr(t)^2} dt
\end{equation}

\noindent
where $p_{\theta}$ and $p_{\phi_1}$ are constants in time. Assuming a DART-like excitation, the linear solutions are those of equations (\ref{LinSol}). Substituting the latter in (\ref{EqTheta}), yields

\begin{equation}\label{EqThetaWithLinSubs}
    \theta(t) = \int \frac{p_{\theta} - p_{\phi_1} - \big(p_{\phi_{2eq,new}} + A_{1p_{\phi 2}}\cos{(\omega_1t)} + A_{2p_{\phi 2}}\cos{(\omega_2t)\big)} }{m\big(r_{eq,new} + A_{1r}\cos{(\omega_1t)} + A_{2r}\cos{(\omega_2t)} \big)^2} dt
\end{equation}

\noindent
where the coefficients $A_{1p_{\phi 2}}$, $A_{2p_{\phi 2}}$ are assumed to be the amplitudes of $\delta p_{\phi_2}(t)$ of equations (\ref{LinVar}), i.e.

\begin{equation}
A_{1p_{\phi 2}} = \frac{(j_{13}j_{31} + \omega_1^2)A_{1r}}{j_{13}j_{21}}, \hspace{0.5cm}
A_{2p_{\phi 2}} = \frac{(j_{13}j_{31} + \omega_2^2)A_{2r}}{j_{13}j_{21}}~~~.
\end{equation}

\noindent
Since the direct evaluation of the integral of equation (\ref{EqThetaWithLinSubs}) proves challenging, we rewrite it as

\begin{equation}\label{EqThetaWithDeltas}
    \theta(t) = \int \frac{p_{\theta} - p_{\phi_1} - \big(p_{\phi_{2eq,new}} + A_{1p_{\phi 2}}\cos{(\omega_1t)} + A_{2p_{\phi 2}}\cos{(\omega_2t)}\big) }{m\big(r_{eq,new} + \delta r(t) \big)^2} dt~~~.
\end{equation}

\noindent
Substituting $\delta r \leftarrow \epsilon \cdot \delta r$, expanding up to first order terms in powers of $\epsilon$ and back substituting the function $\delta r(t)$, renders the integral analytically computable. By separating the polynomial and trigonometric terms, we obtain

\begin{equation}
\begin{split}
    \theta(t;\beta) & = \theta(0) + \bigg(
    \frac{ A_{1r}A_{1p_{\phi 2}} + A_{2r}A_{2p_{\phi 2}} +
           mr_{eq,new}^3\dot{\theta}_{eq,new} }{mr_{eq,new}^3}\bigg)t \\
    & + \frac{-2A_{1r}mr_{eq,new}\dot{\theta}_{eq,new} - A_{1p_{\phi 2}} }{mr_{eq,new}^2\omega_1}\sin{(\omega_1t)} \\
    & + \frac{-2A_{2r}mr_{eq,new}\dot{\theta}_{eq,new} - A_{2p_{\phi 2}} }{mr_{eq,new}^2\omega_2}\sin{(\omega_2t)} \\
    & + \frac{A_{1r}A_{1p_{\phi 2}}}{2mr_{eq,new}^3\omega_1}\sin{(2\omega_1t)} \\
    & + \frac{A_{2r}A_{2p_{\phi 2}}}{2mr_{eq,new}^3\omega_2}\sin{(2\omega_2t)} \\
    & + \frac{A_{1r}A_{2p_{\phi 2}} + A_{2r}A_{1p_{\phi 2}}}{mr_{eq,new}^3(\omega_1-\omega_2)}\sin{\big[(\omega_1-\omega_2)t\big]} \\
    & + \frac{A_{1r}A_{2p_{\phi 2}} + A_{2r}A_{1p_{\phi 2}}}{mr_{eq,new}^3(\omega_1+\omega_2)}\sin{\big[(\omega_1+\omega_2)t\big]}~~~.
\end{split}
\end{equation}

\noindent
The coefficient of each of the above terms corresponds to $A_{j\theta}$ of equation (\ref{PapThetat}). On the other hand $\dot{\phi}_1$, obeys the equation

\begin{equation}
    \dot{\phi}_1 = \frac{p_{\phi_1}}{I_{1z}} -\dot{\theta}~~~.
\end{equation}

\noindent
Direct integration yields

\begin{equation}
    \phi_1(t;\beta) = \phi_1(0) + \frac{p_{\phi_1}}{I_{1z}}t - \theta(t;\beta)~~~.
\end{equation}


\section{Full form of the terms $Z_0$, $h_1$ and $h_2$}\label{app:z0h1h2}

\begin{equation}
\begin{split}
    Z_0  & = \frac{1}{2}\bigg( \nu_1^2I_{1z} + \frac{G\big( M_1(2I_{2x}-I_{2y} - M_2r^{\ast 2}) + M_2(-I_{1z}+I_{2z}+I_s) \big)}{r^{\ast 3}} \bigg) \\
    & + \nu_\theta^{\ast}\delta p_\theta + (\nu_1 - \nu_\theta^{\ast}) \delta p_{\phi_1} + \frac{1}{2m}\delta p_r^2 +
    \bigg( \frac{1}{2I_{2z}} + \frac{1}{2mr^{\ast 2}} \bigg) \delta p_{\phi_2}^2 +
    \frac{2\nu_\theta^{\ast}}{r^{\ast 2}}\delta r \delta p_{\phi_2} \\
    & +\frac{G\big(6M_2(I_s-I_{1z}) + 6M_1(2I_{2x}-I_{2y}-I_{2z}) + M_1M_2r^{\ast 2}\big)}{2r^{\ast 5}} \delta 
    r^2 \\ 
    & - \frac{3GM_1(I_{2x}-I_{2y})}{2 r^{\ast 3}}\delta \phi_2^2
\end{split}
\end{equation}

\begin{equation}
\begin{split}
    h_1 & = \frac{\delta p_{\phi_1} - \delta p_{\theta}}{mr^{\ast 2}}\delta p_{\phi_2}
            -\frac{3\nu_\theta^{\ast}}{r^{\ast 3}}\delta p_{\phi_2} \delta r^2 \\
            & + \frac{G\big( 5M_2(I_{1z} - I_s) + 5M_1(-2I_{2x} + I_{2y} + I_{2z}) - M_1M_2r^{\ast 2} \big)}{r^{\ast 6}}\delta r^3 \\
        &   + \bigg( \frac{3G\big( M_2(I_{1z} - I_s) + M_1(-2I_{2x} + I_{2y} + I_{2z}) \big)}{2 r^{\ast 4}} + \frac{2\nu_\theta^{\ast} (\delta p_{\phi_1} - \delta p_\theta)}{r^{\ast 2}} \bigg) \delta r \\
        & - \frac{1}{mr^{\ast 3}}\delta r \delta p_{\phi_2}^2 + \frac{9GM_1(I_{2x}-I_{2y})}{2 r^{\ast 4}}\delta r \delta \phi_2^2
\end{split}
\end{equation}

\begin{equation}
\begin{split}
 h_2 & = \frac{I_{1z}(\delta p_\theta - \delta p_{\phi_1})^2 + mr^{\ast 2}\delta p_{\phi_1}^2}{2I_{1z}mr^{\ast 2}} +
         \frac{2(\delta p_\theta - \delta p_{\phi_1})}{mr^{\ast 3}} \delta r \delta p_{\phi_2} +
         \frac{4\nu_\theta^{\ast}}{r^{\ast 4}}\delta r^3 \delta p_{\phi_2} \\
     & + \frac{3G\big( 5M_2(I_s - I_{1z}) + 5M_1(2I_{2x} - I_{2y} - I_{2z}) + M_1M_2r^{\ast 2} \big)}{2r^{\ast 7}}\delta r^4  \\
     & + \frac{GM_1(I_{2x}-I_{2y})}{2r^{\ast 3}}\delta \phi_2^4 +
     \frac{3\nu_\theta^{\ast} (\delta p_\theta - \delta p_{\phi_1})}{r^{\ast 3}}\delta r^2 \\
     & + \frac{3}{2mr^{\ast 4}}\delta r^2 \delta p_{\phi_2}^2 -
     \frac{9GM_1(I_{2x}-I_{2y})}{r^{\ast 5}}\delta r^2 \delta \phi_2^2
\end{split}
\end{equation}


\section{Conversion to Birkhoff complex canonical variables $(Q_1, Q_2, P_1, P_2)$}\label{app:QPtrans}

We write down the Hamilton's equations that correspond to the term $Z_0$ of equation (\ref{Hexp})

\begin{equation}
\begin{gathered}
    \dot{\delta r} =    \frac{\partial Z_0}{\partial \delta p_r} \hspace{1.1cm}
    \dot{\delta p}_r = -\frac{\partial Z_0}{\partial \delta r} \\
    \dot{\delta \phi}_2 = \frac{\partial Z_0}{\partial \delta p_{\phi_2}} \hspace{1cm}
    \dot{\delta p}_{\phi_2} = -\frac{\partial Z_0}{\partial \delta \phi_2}
\end{gathered}
\end{equation}

\noindent
and then we evaluate the corresponding Jacobian matrix

\begin{equation}\label{DiagJac}
\boldsymbol{K} =
\begin{bmatrix}
      0    &   0    & k_{13} &   0    \\
    k_{21} &   0    &   0    & k_{24} \\
    k_{31} &   0    &   0    & -k_{21} \\
      0    & k_{42} &   0    &   0
\end{bmatrix}
\end{equation}

\noindent
where

\begin{equation}
   k_{13} = \frac{1}{m}
\end{equation}

\begin{equation}
   k_{21} = \frac{2\nu_\theta^{\ast}}{r^{\ast 2}}
\end{equation}

\begin{equation}
   k_{24} = \frac{1}{I_{2z}} + \frac{1}{mr^{\ast 2}}
\end{equation}

\begin{equation}
   k_{31} = \frac{G \big(6M_2(I_{1z} - I_s) + 6 M_1(-2I_{2x} + I_{2y} + I_{2z}) - M_1M_2 r^{\ast2}\big)}{r^{\ast5}}
\end{equation}

\begin{equation}
   k_{42} = \frac{3GM_1(I_{2x} - I_{2y})}{r^{\ast3}}~~~.
\end{equation}

\noindent
$\boldsymbol{K}$ resembles the Jacobian $\boldsymbol{J}$ of equation (\ref{LinJac}), but in fact $\boldsymbol{K}$ is evaluated for the two-spheres equilibrium point of equation (\ref{TwoSpheresEq}). The eigenvalues of $\boldsymbol{K}$ are obtained from the characteristic polynomial

\begin{equation}\label{PolKep}
\Pi(\lambda) = \lambda ^4 + \zeta_2\lambda ^2 + \zeta_0
\end{equation}

\begin{equation}\label{zeta1}
\begin{split}
    \zeta_2 & = \frac{G M_1 r^{\ast2}}{I_{2z} m r^{\ast5}} \\
    & \times \bigg[ (-3 I_{2x} m + 3 I_{2y} m + I_{2z} M_2) \\
    & - 3G I_{2z} \big(2 M_2 (I_{1z} - I_s) +
    M_1 (-3 I_{2x} + I_{2y} + 2 I_{2z})\big) \bigg]
\end{split}
\end{equation}

\begin{equation}\label{zeta2}
\begin{split}
    \zeta_0 & = \frac{3G^2 M_1 (I_{2x} - I_{2y})}{I_{2z} m^2 r^{\ast10}} \\
    & \times \bigg[4M_1M_2I_{2z}r^{\ast 2} + (I_{2z} + m r^{\ast2})\big(6M_2(I_{1z} - I_s) \\
    & + 6M_1(-2I_{2x} + I_{2y} + I_{2z}) - M_1M_2 r^{\ast2}\big) \bigg]
\end{split}
\end{equation}

\noindent
The roots are $\lambda_{1,2} = \mp i\omega_{1k}$, $\lambda_{3,4} = \mp i\omega_{2k}$ and $\omega_{1k}, \omega_{2k}$ are the kernel frequencies.

\begin{equation}\label{KepOmegas}
    \omega_{1k,2k} = \frac{\sqrt{\zeta_2 \pm \sqrt{\zeta_2 ^2-4 \zeta_0 } }}{\sqrt{2}}~~~.
\end{equation}

\noindent
The corresponding eigenvectors are

\begin{equation}\label{DiagJacVecs}
\boldsymbol{V} =
\begin{bmatrix}
    \vec{\mathrm{v}}_1   \\
    \vec{\mathrm{v}}_2   \\
    \vec{\mathrm{v}}_3   \\
    \vec{\mathrm{v}}_4 
\end{bmatrix} =
\begin{bmatrix}
      f(\lambda_1) & g(\lambda_1) & h(\lambda_1) & 1 \\
      f(\lambda_2) & g(\lambda_2) & h(\lambda_2) & 1 \\
      f(\lambda_3) & g(\lambda_3) & h(\lambda_3) & 1 \\
      f(\lambda_4) & g(\lambda_4) & h(\lambda_4) & 1
\end{bmatrix}
\end{equation}

\noindent
where

\begin{equation}
\begin{gathered}
f(\lambda) = \frac{k_{13}k_{21}}{k_{13}k_{31} - \lambda^2} \\
g(\lambda) = \frac{-k_{13}(k_{21}^2+k_{24}k_{31})+k_{24}\lambda^2}{-k_{13}k_{31}\lambda + \lambda^3} \\
h(\lambda) = \frac{k_{21}\lambda}{k_{13}k_{31} - \lambda^2}~~~.
\end{gathered}
\end{equation}

\noindent
The transformation between the variables $(\delta r, \delta \phi_2, \delta p_r, \delta p_{\phi_2})$ and $(Q_1,Q_2,P_1,P_2)$ is canonical if the symplectic condition is satisfied, i.e.

\begin{equation}\label{Symplecticity}
    \boldsymbol{M}^T \boldsymbol{S} \boldsymbol{M} = \boldsymbol{S}
\end{equation}

\noindent
where $\boldsymbol{M}$ is the $4\times4$ matrix

\begin{equation}\label{Mmatrix}
\boldsymbol{M} =
\begin{bmatrix}
      P_1f(\lambda_1) & P_2f(\lambda_3) & -P_1f(\lambda_2) & -P_2f(\lambda_4) \\
      P_1g(\lambda_1) & P_2g(\lambda_3) & -P_1g(\lambda_2) & -P_2g(\lambda_4) \\
      P_1h(\lambda_1) & P_2h(\lambda_3) & -P_1h(\lambda_2) & -P_2h(\lambda_4) \\
      P_1             & P_2             & -P_1             & -P_2
\end{bmatrix}
\end{equation}

\noindent
and $\boldsymbol{S}$ is the $4\times4$ symplectic matrix

\begin{equation}
\boldsymbol{S} =
\begin{bmatrix}
       0 &  0 & 1 & 0 \\
       0 &  0 & 0 & 1 \\
      -1 &  0 & 0 & 0 \\
       0 & -1 & 0 & 0
\end{bmatrix}~~~.
\end{equation}

\noindent
By substituting (\ref{Mmatrix}) into (\ref{Symplecticity}) and recalling that $\lambda_2=-\lambda_1$, $\lambda_4=-\lambda_3$, $f(-\lambda) = f(\lambda)$, $g(-\lambda) = -g(\lambda)$ and $h(-\lambda) = -h(\lambda)$, we obtain the two equations

\begin{equation}\label{NonTrivialQPEqs}
2\big(g(\lambda_1) - f(\lambda_1)h(\lambda_1)\big)P_1^2 + 1 = 0, \hspace{0.6cm}
2\big(g(\lambda_3) - f(\lambda_3)h(\lambda_3)\big)P_2^2 + 1 = 0~~~.
\end{equation}

\noindent
We choose one of the corresponding set of roots

\begin{equation}\label{NonTrivialQPEqsRoots}
P_1 = \frac{-i}{u(\lambda_1)}, \hspace{0.5cm}
P_2 = \frac{-i}{u(\lambda_3)}, \hspace{0.3cm} \text{where } \hspace{0.2cm}
u(\lambda) = \sqrt{2g(\lambda) - 2f(\lambda)h(\lambda)}
\end{equation}

\noindent
Substituting (\ref{NonTrivialQPEqsRoots}) back at (\ref{Mmatrix}), yields

\begin{equation}
\boldsymbol{M} \equiv
\begin{bmatrix}
m_{11} & m_{12} & m_{13} & m_{14} \\
m_{21} & m_{22} & m_{23} & m_{24} \\
m_{31} & m_{32} & m_{33} & m_{34} \\
m_{41} & m_{42} & m_{43} & m_{44}
\end{bmatrix} = 
\begin{bmatrix}
-\frac{\displaystyle if(\lambda_1)}{\displaystyle u(\lambda_1)} &
-\frac{\displaystyle if(\lambda_3)}{\displaystyle u(\lambda_3)} &
 \frac{\displaystyle if(\lambda_1)}{\displaystyle u(\lambda_1)} &
 \frac{\displaystyle if(\lambda_3)}{\displaystyle u(\lambda_3)} \\[10pt]
-\frac{\displaystyle ig(\lambda_1)}{\displaystyle u(\lambda_1)} &
-\frac{\displaystyle ig(\lambda_3)}{\displaystyle u(\lambda_3)} &
-\frac{\displaystyle ig(\lambda_1)}{\displaystyle u(\lambda_1)} &
-\frac{\displaystyle ig(\lambda_3)}{\displaystyle u(\lambda_3)} \\[10pt]
-\frac{\displaystyle ih(\lambda_1)}{\displaystyle u(\lambda_1)} &
-\frac{\displaystyle ih(\lambda_3)}{\displaystyle u(\lambda_3)} &
-\frac{\displaystyle ih(\lambda_1)}{\displaystyle u(\lambda_1)} &
-\frac{\displaystyle ih(\lambda_3)}{\displaystyle u(\lambda_3)} \\[10pt]
-\frac{\displaystyle i}{\displaystyle u(\lambda_1)} &
-\frac{\displaystyle i}{\displaystyle u(\lambda_3)} &
 \frac{\displaystyle i}{\displaystyle u(\lambda_1)} &
 \frac{\displaystyle i}{\displaystyle u(\lambda_3)} \\
\end{bmatrix}
\end{equation}

\noindent
and the inverse matrix $\boldsymbol{M}^{-1}$ is

\begin{equation}
\boldsymbol{M}^{-1} =
\begin{bmatrix}
 \frac{\displaystyle iu(\lambda_1)}{\displaystyle q_1} &
 \frac{\displaystyle ih(\lambda_3)u(\lambda_1)}{\displaystyle q_2} &
 \frac{\displaystyle ig(\lambda_3)u(\lambda_1)}{\displaystyle q_3} &
-\frac{\displaystyle if(\lambda_3)u(\lambda_1)}{\displaystyle q_4} \\[10pt]
-\frac{\displaystyle iu(\lambda_3)}{\displaystyle q_1} &
 \frac{\displaystyle ih(\lambda_1)u(\lambda_3)}{\displaystyle q_3} &
 \frac{\displaystyle ig(\lambda_1)u(\lambda_3)}{\displaystyle q_2} &
 \frac{\displaystyle if(\lambda_1)u(\lambda_3)}{\displaystyle q_4} \\[10pt]
-\frac{\displaystyle iu(\lambda_1)}{\displaystyle q_1} &
 \frac{\displaystyle ih(\lambda_3)u(\lambda_1)}{\displaystyle q_2} &
 \frac{\displaystyle ig(\lambda_3)u(\lambda_1)}{\displaystyle q_3} &
 \frac{\displaystyle if(\lambda_3)u(\lambda_1)}{\displaystyle q_4} \\[10pt]
 \frac{\displaystyle iu(\lambda_3)}{\displaystyle q_1} &
 \frac{\displaystyle ih(\lambda_1)u(\lambda_3)}{\displaystyle q_3} &
 \frac{\displaystyle ig(\lambda_1)u(\lambda_3)}{\displaystyle q_2} &
-\frac{\displaystyle if(\lambda_1)u(\lambda_3)}{\displaystyle q_4} \\[10pt]
\end{bmatrix}
\end{equation}
\vspace{0.2cm}

where 
\begin{align*}
q_1 &= \displaystyle 2\big(f(\lambda_1) - f(\lambda_3)\big) \\
q_2 &= \displaystyle 2\big(-g(\lambda_3)h(\lambda_1) + g(\lambda_1)h(\lambda_3)\big) \\
q_3 &= \displaystyle 2\big( g(\lambda_3)h(\lambda_1) - g(\lambda_1)h(\lambda_3)\big) \\
q_4 &= \displaystyle 2\big( f(\lambda_1) - f(\lambda_3)\big)
\end{align*}
\noindent
Ultimately, $\boldsymbol{M}$ and $\boldsymbol{M}^{-1}$ set the transformations (forward and inverse) between
$(\delta r, \delta \phi_2, \delta p_r, \delta p_{\phi_2})$ and $(Q_1,Q_2,P_1,P_2)$ as 
functions of the physical parameters and $r^{\ast}$.


\section{Evaluation of the normal form Hamiltonian}\label{app:EvalNormalForm}

\noindent
The Birkhoff normalization process defines a sequence of $N$ consecutive near-to-identity canonical transformations to the variables $\boldsymbol{\mathcal{F}} = (Q_1,Q_2,P_1,P_2)$, namely

\begin{equation}\label{QPTOQP4}
\boldsymbol{\mathcal{F}} \equiv \boldsymbol{\mathcal{F}}^{(0)} \rightarrow \boldsymbol{\mathcal{F}}^{(1)} \rightarrow \hspace{0.1cm} ... \rightarrow \hspace{0.1cm}  \boldsymbol{\mathcal{F}}^{(N)}~~~,
\end{equation}

\noindent
such that, in the variables $\boldsymbol{\mathcal{F}}^{(N)}$, the Hamiltonian, resumes the form

\begin{equation}\label{HQP4}
    H^{(N)} = \bigg( Z_0 + \epsilon Z_1 + ... + \epsilon^N Z_N \bigg) +
    \sum_{j=N+1}^{\infty}\epsilon^j H_j^{(N)}~~~.
\end{equation}

\noindent
The quantity

\begin{equation}\label{NormalForm}
    Z^{(N)} = Z_0 + \epsilon Z_1 + ... + \epsilon^N Z_N~~~,
\end{equation}

\noindent
is in Birkhoff normal form, implying that $Z^{(N)}$ is integrable. The quantity

\begin{equation}\label{HQPRemainder}
    R^{(N)} = \sum_{j=N+1}^{\infty}\epsilon^j H_j^{(N)}~~~,
\end{equation}

\noindent
is called the remainder function. Various theoretical arguments (see e.g. \cite{Efthymiopoulos2012}) imply that the remainder yields, in general, a small correction which can be ignored in the equations of motion. The functional form of the transformations (\ref{QPTOQP4}) is obtained by the method of composition of Lie series (see Appendix \ref{app:NF4order}). Up to the $N_{th}$ order we get

\begin{equation}\label{LieTrans}
  \boldsymbol{\mathcal{F}} \equiv \boldsymbol{\mathcal{F}}^{(0)} = \exp{(\mathcal{L}_{\chi_N})}\exp{(\mathcal{L}_{\chi_{N-1}})} \cdots 
  \exp{(\mathcal{L}_{\chi_1})}\boldsymbol{\mathcal{F}}^{(N)}
\end{equation}

\noindent
which expresses the variables $\boldsymbol{\mathcal{F}}^{(0)}$ in terms of $\boldsymbol{\mathcal{F}}^{(N)}$. The operator $\exp{(\mathcal{L}_{\chi})}$, acting on an arbitrary function $f$, is defined as

\begin{equation}
  \exp{(\mathcal{L}_{\chi})}f = f +
                                \mathcal{L}_{\chi}f +
                                \frac{\mathcal{L}_{\chi}^2f}{2} +
                                \frac{\mathcal{L}_{\chi}^3f}{6} + \cdots~~~,
\end{equation}

\noindent
where $\mathcal{L}_{\chi}$ denotes the Poisson bracket operator, $\mathcal{L}_{\chi}f = \{ f, \chi \}$. The functions $\chi_1, ..., \chi_N$ are called Lie generating functions and at each normalization step, they are evaluated through the solution of the homological equation

\begin{equation}
  \{Z_0, \chi_{n+1}\} + \epsilon^{n+1}h_{n+1}^{(n)} = 0~~~,
\end{equation}

\noindent
where $Z_0$ is the kernel of the normal form, and $h_{n+1}^{(n)}$ contains all the terms of book-keeping order $n+1$ in the Hamiltonian $H^{(n)}$ obtained after $n$ normalization steps, for which the Poisson bracket with $Z_0$ is non-zero. After performing $N$ steps of the normalization algorithm, we obtain the normal form Hamiltonian (\ref{ExpandedZPol}). Note that during the normalization steps, all terms of book-keeping order greater than $N$ are truncated $(O(\epsilon^ {q > N}) = 0)$.

In order to map the solution (\ref{FinSolQP4}) into the original variables $\big(\delta r(t), \delta \phi_2(t), \delta p_r(t), \delta p_{\phi_2}(t)\big)$ and $\big( \theta(t), \phi_1(t) \big)$, we implement the following two steps:

\noindent
1) Assuming an initial disturbance $\big(\delta r(0)=0$, $\delta \phi_2(0)=0$, $\delta p_r(0)=0$, $\delta p_\theta = p_{\theta_{imp}} - p_\theta^{\ast}$, $\delta p_{\phi_1} = 0$, $\delta p_{\phi_2}(0) = p_{\phi_{2imp}} - p_{\phi_2}^{\ast} \big)$, we use the inverse transformation (\ref{delta2QP}) to compute the initial conditions $\boldsymbol{\mathcal{F}}^{(0)}(0)$, and then the inverse of the Lie series (\ref{LieTrans}) (see Appendix \ref{app:NF4order})

\begin{equation}\label{InvLieTrans}
\boldsymbol{\mathcal{F}}^{(N)} = \exp{(\mathcal{-L}_{\chi_1})}\cdots\exp{(\mathcal{-L}_{\chi_{N-1}})}
  \exp{(\mathcal{-L}_{\chi_N})}\boldsymbol{\mathcal{F}}^{(0)}
\end{equation}

\noindent
in order to compute the initial conditions $\boldsymbol{\mathcal{F}}^{(4)}(0)$. By doing so, we can express $\omega_{1c}, \omega_{2c}, \omega_{\theta c}, \omega_{\phi_1 c}$ as functions of $\beta$, as noted in equation (\ref{nu12}). 2) We transform the whole solution in time back to the original variables. This is accomplished by utilizing the Lie series (\ref{LieTrans}) to convert equation (\ref{FinSolQP4}) to $\boldsymbol{\mathcal{F}}^{(0)}(t)$ and $\big( \theta(t), \phi_1(t) \big)$. Finally the transformation (\ref{delta2QP}) yields the desired functions $\big(\delta r(t), \delta \phi_2(t), \delta p_r(t), \delta p_{\phi_2}(t)\big)$.


\section{Composition of Lie series up to fourth order}\label{app:NF4order}

\begin{equation}
\begin{split}
e^{\mathcal{L}_{\chi_4}}e^{\mathcal{L}_{\chi_3}}e^{\mathcal{L}_{\chi_2}}e^{\mathcal{L}_{\chi_1}}f & =
\big(1 +
\mathcal{L}_{\chi_1} +
\frac{\mathcal{L}_{\chi_1}^2}{2} +
\frac{\mathcal{L}_{\chi_1}^3}{6} +
\frac{\mathcal{L}_{\chi_1}^4}{24} +
\mathcal{L}_{\chi_2} +
\mathcal{L}_{\chi_2}\mathcal{L}_{\chi_1} \\
& + \mathcal{L}_{\chi_2}\frac{\mathcal{L}_{\chi_1}^2}{2} +
\frac{\mathcal{L}_{\chi_2}^2}{2} +
\mathcal{L}_{\chi_3} +
\mathcal{L}_{\chi_3}\mathcal{L}_{\chi_1} +
\mathcal{L}_{\chi_4}\big)f \\
& = f +
\{ f,\chi_1 \} +
\frac{1}{2}\{ \{ f,\chi_1 \}, \chi_1 \} +
\frac{1}{6}\{ \{ \{ f,\chi_1 \}, \chi_1 \}, \chi_1 \} \\
& + \frac{1}{24}\{ \{ \{ \{ f,\chi_1 \}, \chi_1 \}, \chi_1 \}, \chi_1 \} +
\{ f,\chi_2 \} +
\{ \{ f,\chi_1 \}, \chi_2 \} \\
& + \frac{1}{2}\{ \{ \{ f,\chi_1 \}, \chi_1 \}, \chi_2 \} +
\frac{1}{2}\{ \{ f,\chi_2 \}, \chi_2 \} +
\{ f,\chi_3 \} \\
& +\{ \{ f,\chi_1 \}, \chi_3 \} +
\{ f,\chi_4 \}
\end{split}
\end{equation}

\begin{equation}
\begin{split}
e^{\mathcal{-L}_{\chi_1}}e^{\mathcal{-L}_{\chi_2}}e^{\mathcal{-L}_{\chi_3}}e^{\mathcal{-L}_{\chi_4}}f & =
\big(1 -
\mathcal{L}_{\chi_1} +
\frac{\mathcal{L}_{\chi_1}^2}{2} -
\frac{\mathcal{L}_{\chi_1}^3}{6} +
\frac{\mathcal{L}_{\chi_1}^4}{24} -
\mathcal{L}_{\chi_2} +
\mathcal{L}_{\chi_1}\mathcal{L}_{\chi_2} \\ 
& - \frac{\mathcal{L}_{\chi_1}^2}{2}\mathcal{L}_{\chi_2} +
\frac{\mathcal{L}_{\chi_2}^2}{2} -
\mathcal{L}_{\chi_3} +
\mathcal{L}_{\chi_1}\mathcal{L}_{\chi_3} -
\mathcal{L}_{\chi_4}\big)f \\
& = f -
\{ f,\chi_1 \} +
\frac{1}{2}\{ \{ f,\chi_1 \}, \chi_1 \} -
\frac{1}{6}\{ \{ \{ f,\chi_1 \}, \chi_1 \}, \chi_1 \} \\
& + \frac{1}{24}\{ \{ \{ \{ f,\chi_1 \}, \chi_1 \}, \chi_1 \}, \chi_1 \} -
\{ f,\chi_2 \} +
\{ \{ f,\chi_2 \}, \chi_1 \} \\ 
& - \frac{1}{2}\{ \{ \{ f,\chi_2 \}, \chi_1 \}, \chi_1 \} +
\frac{1}{2}\{ \{ f,\chi_2 \}, \chi_2 \} -
\{ f,\chi_3 \} \\
& +\{ \{ f,\chi_3 \}, \chi_1 \} -
\{ f,\chi_4 \}
\end{split}
\end{equation}

\end{appendices}


\bibliography{sn-bibliography}

\end{document}